\documentclass[runningheads,a4paper]{llncs}

\usepackage{rotating}

\usepackage{extarrows}
 \usepackage{amsmath, amssymb, latexsym}
 \usepackage{graphicx}
 \usepackage[boxed]{algorithm}
 \usepackage{color}
 \usepackage{caption}
\usepackage{algorithm}

\usepackage{amsthm}
\usepackage{multirow}

 \usepackage{psfrag}

   \newtheorem*{citetheorem}{Theorem}

  \newenvironment{Proof}{\noindent {\bf Proof:\,}}{\hfill$\;\;\Box$ \\}

  \def	\Longlonghookrightarrow{\lhook\joinrel\relbar\joinrel\relbar\joinrel\relbar\joinrel\relbar\joinrel\rightarrow}
  \newcommand{\smrule}[1]{\stackrel{#1}{\Longlonghookrightarrow}}
  
  \def\by#1{\mathop{{\hbox{\setbox0=\hbox{$\scriptstyle{#1\quad}$}{$%
  					\mathrel{\mathop{\setbox1=\hbox to \wd0{\rightarrowfill}\ht1=3pt\dp1=-2pt\box1}\limits^{#1}}%
  					$}}}}}
  					
  \def\tr#1{\mathop{{\hbox{\setbox0=\hbox{$\scriptstyle{#1\quad}$}{$%
  					\mathrel{\mathop{\setbox1=\hbox to \wd0{\rightarrow}\ht1=3pt\dp1=-2pt\box1}\limits^{#1}}%
  					$}}}}}

\usepackage{url}
\urldef{\mailsa}\path| |

\begin{document}

\mainmatter  

\title{LTL Model Checking of Self Modifying Code}

\titlerunning{LTL Model Checking of Self Modifying Code}

%
%
\author{Tayssir Touili$^1$
\and Xin Ye$^1$$^2$}
\authorrunning{Lecture Notes in Computer Science: Authors' Instructions}

\institute{$^1$CNRS,LIPN and University Paris 13\\
$^2$East China Normal University, Shanghai, China\\
\mailsa\\
}
%
%

\toctitle{Lecture Notes in Computer Science}
\tocauthor{Authors' Instructions}
\maketitle

\begin{abstract}
Self modifying code is code that can modify its own instructions
 during the execution of the program. It is extensively used by malware writers to obfuscate their malicious code.
Thus, analysing self modifying code is nowadays a big challenge.
In this paper,  we consider the LTL model-checking problem of self modifying code.
We model such programs using self-modifying pushdown systems (SM-PDS),  an extension of  pushdown systems that can modify its own set of transitions during execution.
We reduce the LTL model-checking  problem to the emptiness problem of self-modifying B\"uchi pushdown systems (SM-BPDS).
We implemented our techniques in a tool that  we successfully applied for the detection of several self-modifying malware.
Our tool was also able to detect several malwares that well-known antiviruses such as BitDefender, Kinsoft, Avira, eScan, Kaspersky, Qihoo-360, Baidu, Avast, and Symantec 
failed  to detect.
\end{abstract}

\section{Introduction}
Binary code presents several complex  aspects that cannot be encountred in source code. One of these aspects is self-modifying code, i.e., code that can modify its own instructions
 during the execution of the program.  Self-modifying code   makes reverse code engineering harder. Thus, it is extensively used to protect software intellectual property.
It is also heavily used by malware writers in order to make their malwares hard to analyse and detect by static analysers and anti-viruses. Thus, it is crucial to be able to analyse 
self-modifying code.

There are several kinds of  self-modifying code. In this work, we consider    self-modifying code  caused by \textbf{self-modifying instructions}.
These kind of instructions treat code as data. This allows them to read and write into code, leading to \textbf{self-modifying instructions}.
 These  self-modifying instructions are usually {\bf mov} instructions, since {\bf mov} allows to access memory and read and write into it.

\begin{figure}[htbp] 
 		\centering
 		\includegraphics[width=0.98\textwidth]{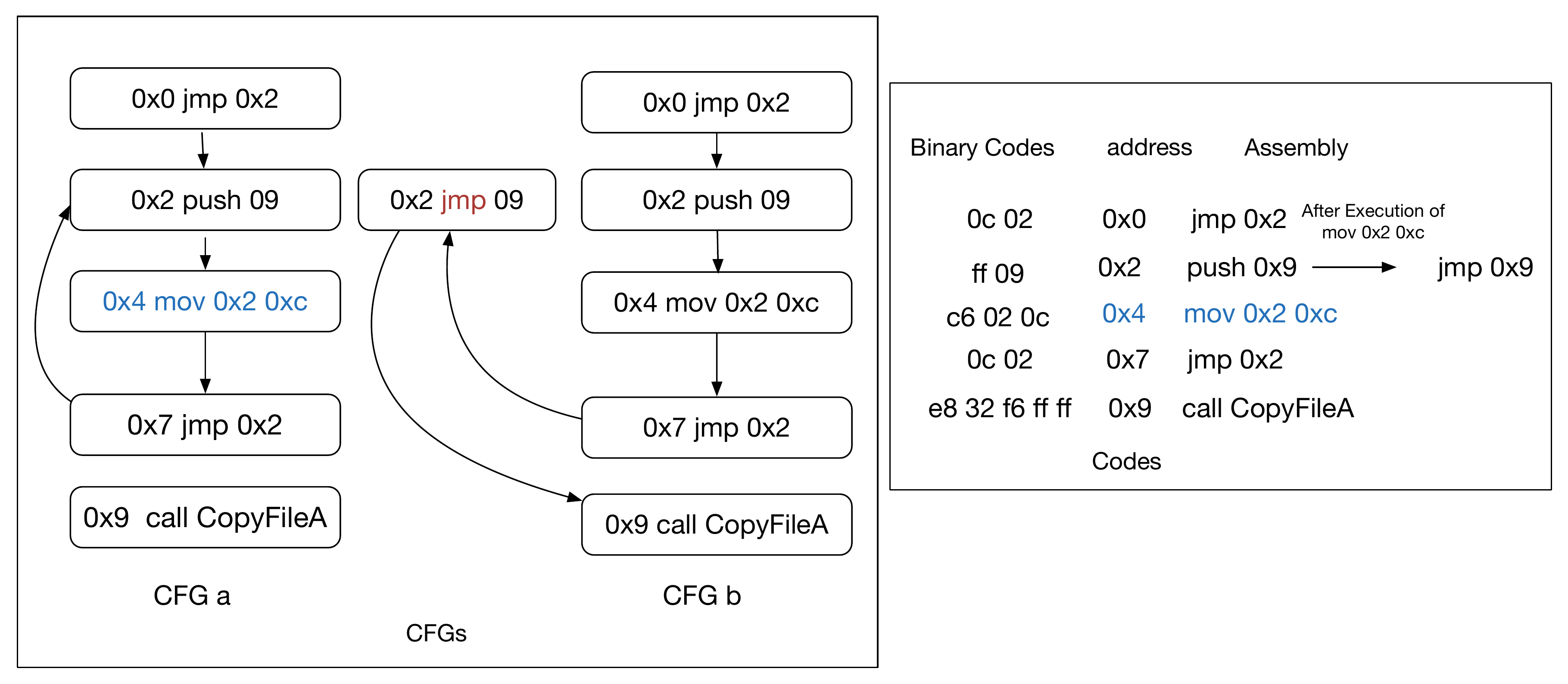}
 		\caption{An Example of a Self-modifying Code}\label{fig:bcc}
 	\end{figure}

Let us consider the example shown in  \figurename{\ref{fig:bcc}}.
For simplicity, the addresses' length   is assumed to be 1 byte.
In the right box, we give,  respectively, the binary code, the addresses of the different instructions, and the  corresponding assembly code, obtained  by translating 
syntactically the binary code at  each address. For example,  {\tt  0c} is the binary code of the jump {\tt  jmp}.
Thus, {\tt  0c 02} is translated to {\tt jmp  0x2} (jump to address  0x2). The second line is translated to  {\tt push 0x9},
since {\tt ff} is the binary code of the instruction {\tt push}.
The third instruction {\tt mov 0x2 0xc} will replace the first byte at address 
{\tt 0x2} by  {\tt 0xc}. Thus, at address {\tt  0x2}, {\tt  ff 09} is replaced by {\tt  0c 09}.
This means the instruction    {\tt  push 0x9} is replaced by the jump instruction {\tt jmp  0x9} (jump to address  0x9), etc.
Therefore, this code  is self-modifying: the {\bf  mov} instruction was able to modify the instructions of the program 
via its ability to read and write the memory.
If we study this code without  looking at the semantics of the  self-modifying instructions, we will  extract from it the Control Flow Graph {\tt CFG a}  that is in the left of the figure,
and we will reach the conclusion that the call to the API function  CopyFileA at address {\tt  0x9} cannot be made.
However, you can see that the correct CFG is the one on the right hand side  {\tt CFG b}, where   the call to the API function  CopyFileA at address {\tt  0x9} 
can be reached. Thus, it is very important to be able to take into account  the semantics of the  self-modifying instructions in binary code.

In this paper, we consider the LTL model-checking problem of self-modifying code. 
To this aim, we use Self-Modifying Pushdown Systems (SM-PDSs) \cite{touili2017reachability} to model self-modifying code.
Indeed, SM-PDSs were shown in \cite{touili2017reachability} to be an adequate model for self-modifying code since they allow to mimic  the program's stack  
while taking into account the self-modifying semantics  of the transitions. This is very important for binary code analysis and malware detection, since malwares are based on calls to 
API functions of the operating system. Thus, antiviruses check the API calls to determine whether a program is malicious or not.
Therefore, to evade from these antiviruses, malware writers try to hide the API calls they make by replacing calls by push and jump instructions.
Thus, to be able to analyse such malwares, it is crucial to be able to analyse the program's stack. Hence the need to a model like pushdown systems and self-modifying pushdown systems for this purpose, since they  allow to mimic the program's stack.

Intuitively, a SM-PDS is a pushdown system (PDS) with self-modifying rules, i.e., with rules that allow to modify the current  set of transitions during execution.
This model was introduced in \cite{touili2017reachability} in order to represent self-modifying code.  In \cite{touili2017reachability}, the authors have proposed algrithms to compute 
finite automata that accept the forward and backward reachability sets of SM-PDSs.
In this work, we  tackle the problem of LTL model-checking of SM-PDSs. Since SM-PDSs are equivalent to PDSs \cite{touili2017reachability}, one possible approach for LTL model checking
 of SM-PDS is to translate the SM-PDS to a standard PDS and then run the LTL model checking algorithm on the equivalent PDS   {\cite{Bouajjani:1997ew,esparza}}. But translation from a SM-PDS to a standard PDS is exponential. Thus, performing 
 	the LTL model checking on the equivalent PDS is not efficient.

 	To overcome this limitation, we propose a {\em direct} LTL model checking algorithm for SM-PDSs. 
 	Our algorithm is based on reducing the LTL model checking problem to the emptiness problem of Self Modifying B\"{u}chi Pushdown Systems (SM-BPDS).
Intuitively, we obtain this SM-BPDS  by taking the  product of the SM-PDS with a B\"{u}chi automaton accepting an  LTL formula $\varphi$. 
Then, we solve the emptiness problem of an SM-BPDS by computing its repeating heads. This computation is based on computing labelled $pre^*$ configurations by applying a saturation procedure
 on  labelled finite automata.

 	We implemented our algorithm in a tool. 
 	Our experiments show that our {\em direct} techniques are much more efficient than translating the SM-PDS to an equivalent PDS and then applying the standard 
 	LTL model checking for PDSs \cite{Bouajjani:1997ew,esparza}.
 	Moreover, we  successfully applied our tool to the analysis of 892 self-modifying malwares. 
Our tool was also able to detect several self-modifying malwares that well-known antiviruses like BitDefender, Kinsoft, Avira, eScan, Kaspersky, Qihoo-360, Baidu, Avast, and Symantec 
were not able to detect.

 	\medskip
\noindent
 	\textbf{Related Work.}
 	Model checking and static analysis approaches have been widely used to analyze binary programs, for instance, 
 	in \cite{BERGERON,Balakrishnan,SINGHf,Kinder,SongT12}.  Temporal Logics were  chosen to describe malicious   {behaviors} in \cite{beaucamps2010behavior,Kinder,SongT12,ST13,NguyenT17}. 
 However, these works cannot deal with self-modifying code.

POMMADE \cite{SongT12,ST13} is a malware detector based on LTL and CTL model-checking of PDSs. STAMAD \cite{dam1,dam2,dam3} is   a malware detector based on PDSs and machine learning.
However, POMMADE and STAMAD cannot deal with self-modifying code. 
 	
 	 Cai et al. \cite{cai2007certified} use  local reasoning and separation logic to describe  self-modifying  code 
 	 and   treat  program code uniformly as regular data structure.   However, \cite{cai2007certified}  requires programs to be manually annotated with invariants. 
 	In \cite{Debray2008}, the authors  propose  a formal semantics for  self-modifying codes, and   use that to represent self-unpacking code. This work only deals with packing  and unpacking  behaviours.
 	Bonfante  et al. \cite{bonfante2009computability} provide an operational semantics for self-modifying programs and  show that they  can be constructively
 	rewritten to a non-modifying program. 
 	However, all these specifications \cite{bonfante2009computability,cai2007certified,Debray2008} are  too abstract  to be used in practice. 
 	
In \cite{anckaert2007model}, the authors propose a new representation of self-modifying code named State Enhanced-Control Flow Graph (SE-CFG). SE-CFG extends 
standard control flow graphs with a new data structure, keeping track of the possible states programs can reach, and with  edges that can be conditional on the state of the target memory location. 
It  is not easy to analyse a binary program only using its SE-CFG, especially that this representation does not allow to take into account the stack of the program.

 	\cite{Blazy2016} propose abstract interpretation techniques to compute  an over-approximation of the set of reachable states
 	of a self-modifying program, where for each control point of the program, an over-approximation of the memory state at this control point is provided.
 	\cite{roundy2010hybrid} combine static and dynamic analysis techniques to analyse self-modifying programs.
 	Unlike our approach, these techniques \cite{Blazy2016,roundy2010hybrid} cannot handle the program's stack. 
 	
 	Unpacking binary code is also considered in \cite{Coogan2009,Kang2007,Royal2006,Debray2008}. These works do not consider self-modifying 
 	{\bf mov} instructions.

 	\medskip
 	{\bf Outline.}
 	The rest of the paper is structured as follows: Section 2 recalls the definition of Self Modifying pushdown systems. LTL model checking and SM-BPDSs are defined in Section 3.
Section 4 solves the emptiness  problem of SM-BPDS. Finally, the experiments are reported in Section 5.

 	\newcommand{\States}{\textbf{States}}
 	\newcommand{\Reg}{\textbf{R}}
 	\newcommand{\Val}{\textbf{Val}}
 	\newcommand{\Mem}{\textbf{Mem}}
 	\newcommand{\EXP}{\textbf{EXP}}
 	\newcommand{\Z}{\mathbb{Z}}
 	\newcommand{\Orac}{\mathcal{O}}

   \section{Self Modifying  Pushdown Systems }
 
   \subsection{Definition}
We recall in this section the definition of Self-modifying Pushdown Systems \cite{touili2017reachability}.
 \begin{definition}
A  Self-modifying Pushdown System (SM-PDS) is a tuple $\mathcal{P}=(P,\Gamma,\Delta,\Delta_c)$, 
 			where $P$ is a finite set of control points, $\Gamma$ is a finite set of stack symbols, 
 			$\Delta \subseteq (P \times \Gamma)\times (P \times \Gamma^*)$ is a finite set of transition rules, and 
 			$\Delta_c \in P \times \Delta \times \Delta \times P$ is a finite set of modifying transition rules.
 			If  $((p, \gamma), (p', w)) \in \Delta$, we also write  $\langle p,\gamma \rangle \hookrightarrow \langle p',w \rangle \in \Delta$. 
 			If  $(p,r_1,r_2,p') \in \Delta_c$,  we also write $p \smrule{(r_1,r_2)} p'\in \Delta_c$.  
A Pushdown System  (PDS) is a SM-PDS where $\Delta_c=\emptyset$.
\end{definition}

Intuitively, a Self-modifying Pushdown System is a Pushdown System   that can dynamically
modify its set of rules during the execution time: rules $\Delta$ are standard PDS transition rules, while rules $\Delta_c$
modify the current set of  transition rules:   $\langle p,\gamma \rangle \hookrightarrow \langle p',w \rangle \in \Delta$  expresses that if the 
SM-PDS is in control point $p$ and has $\gamma$
on top of its stack, then it can move to control point $p'$, pop $\gamma$ and push $w$ onto the stack, while
$p \smrule{(r_1,r_2)} p'\in \Delta_c$ expresses that when the PDS is in control point $p$, then it can move to  control point $p'$,
remove the rule $r_1$ from its current set of transition rules, and add the rule $r_2$. 

Formally, a configuration of a SM-PDS is a 
tuple $c=(\langle p,w\rangle,\theta)$ where $p\in P$ is the control point, $w\in\Gamma^*$ is the stack content, 
and $\theta\subseteq\Delta\cup\Delta_c$ is the current set of transition rules of the SM-PDS. $\theta$ is called the current {\em phase} of the SM-PDS. When the SM-PDS is a PDS, i.e., when $\Delta_c=\emptyset$, a configuration is a tuple $c=(\langle p,w\rangle,\Delta)$, since there 
is no changing rule, so there is only one possible phase. In this case, we can also write   $c=\langle p,w\rangle$.
Let $\mathcal{C}$ be the set of configurations of a SM-PDS.
A SM-PDS defines a transition relation $\Rightarrow_{\mathcal{P}}$ between configurations as follows: Let $c=(\langle p,w\rangle,\theta)$ be a configuration,
and let $r$ be a rule in $\theta$,  then:

\begin{enumerate}
	\item if $r\in\Delta_c$ is of the form $r= p \smrule{(r_1,r_2)} p'$, such that $r_1 \in \theta$, then 
	$(\langle p, w \rangle, \theta) \Rightarrow_{\mathcal{P}} (\langle p', w \rangle,\theta')$, where 
	$\theta'=(\theta \setminus \{ r_1\}) \cup\{ r_2\}$. In other words, the transition rule $r$ updates the current set of transition rules $\theta$ by removing 
	$r_1$  from it and adding  $r_2$ to it.
	
	\item if $r\in\Delta$ is of the form $r=\langle p,\gamma \rangle \hookrightarrow \langle p',w'\rangle \in \Delta$, then 
	$(\langle p, \gamma w \rangle, \theta) \Rightarrow_{\mathcal{P}} (\langle p',w'w \rangle,\theta)$. In other words, the transition rule $r$ moves  the control point from $p$ to $p'$, 
	pops $\gamma$ from the stack and pushes $w'$ onto the stack. This transition keeps the current set of transition rules $\theta$ unchanged.
\end{enumerate}

Let  $\Rightarrow_{\mathcal{P}}^*$ be the transitive, reflexive closure of $\Rightarrow_{\mathcal{P}}$ and 
  $\Rightarrow_{\mathcal{P}}^+$ be its  transitive  closure.
 An execution (a run)  of $\mathcal{P}$ is a sequence of configurations $\pi=c_0c_1...$  s.t.  
$c_i \Rightarrow_{\mathcal{P}} c_{i+1}$ for every $i\geq 0$.
Given a configuration $c$, the set of immediate predecessors (resp. successors)
of $c$ is $pre_{\mathcal{P}}(c)=\{c'\in \mathcal{C} \; : \; c' \Rightarrow_{\mathcal{P}} c\}$
(resp. $post_{\mathcal{P}}(c)=\{c'\in \mathcal{C} \; : \; c \Rightarrow_{\mathcal{P}} c'\}$). These notations
can be generalized straightforwardly to sets of configurations.
Let $pre_\mathcal{P}^*$ (resp. $post_\mathcal{P}^*$) denote the reflexive-transitive closure of $pre_\mathcal{P}$ (resp.
$post_\mathcal{P}$). We remove the subscript $\mathcal{P}$ when it is clear from the context.

\medskip
We suppose w.l.o.g. that rules in $\Delta$ are of the form $\langle p,\gamma \rangle \hookrightarrow \langle p',w \rangle$ such that 
$|w|\le 2$, and that the self-modifying rules $r=p \smrule{(r_1,r_2)} p'$ in $\Delta_c$ are such that $r\neq r_1$. 
Note that this is not a restriction, since for a given SM-PDS, one can compute an equivalent SM-PDS that satisfies these conditions \cite{touili2017reachability} .

\subsection{SM-PDS vs. PDS}
\label{versPDS}

Let $\mathcal{P}=(P,\Gamma,\Delta,\Delta_c)$ be a SM-PDS. 
It was shown in \cite{touili2017reachability} that:

\begin{enumerate}
\item  $\mathcal{P}$ can be described by an equivalent pushdown system (PDS). Indeed, since the number of phases is finite, we can encode phases
in the control point of the  PDS.  However, this translation is not efficient since the  number of control points of the equivalent PDS is    
$|P|\cdot 2^{\mathcal{O}(|\Delta| +|\Delta_c|)}$.

\item  $\mathcal{P}$ can also be described by an equivalent Symbolic pushdown system \cite{Schwoon:2007vs}, where each SM-PDS rule is represented by a {\em single, symbolic}
transition, where the different values of the phases are encoded in a symbolic way using relations between phases. 
This translation is not efficient neither since the  size   of the relations used in the symbolic transitions is
	$2^{\mathcal{O}(|\Delta|+|\Delta_c|)}$.

\end{enumerate}

\subsection{From Self-modifying Code to SM-PDS }
\label{model}

It is shown in \cite{touili2017reachability}  how to describe a self-modifying binary code using a SM-PDS.
The basic idea is that  the control locations of the SM-PDS store the control points of the binary program and
the stack  mimics the program's stack.  Our translation relies on the  disassembler Jakstab \cite{jakstab}
to disassemble binary code, construct the control flow graph (CFG), determine indirect jumps,
compute the possible values of used variables,  registers and the memory locations at each control point of program. After getting the control flow graph whose edges are equipped with disassembled instructions, we translate the CFG into a SM-PDS as described in \cite{touili2017reachability}. The non self-modifying instructions  of the program define the rules $\Delta$ of the SM-PDS (which are standard PDS rules),
and can be obtained following the translation of \cite{SongT12} that models non self-modifying instructions  of the program by a PDS.
Self-modifying instructions are represented using self-modifying transitions  $\Delta_c$ of the  SM-PDS.  
For more details, we refer the reader to  \cite{touili2017reachability}.

 \section{ LTL Model-Checking of SM-PDSs}

\subsection{The linear-time temporal logic LTL}

Let $At$ be a finite set of atomic propositions.    LTL formulas  are defined as follows (where $A\in At$):
\begin{equation*}
	\varphi:=~A~|~\neg \varphi ~|~   {\varphi_1 \vee \varphi_2} |~X \varphi ~|~ \varphi_1 U \varphi_2
\end{equation*}

Formulae are interpreted on infinite words  over $2^{At}$.  Let $\omega=\omega^0 \omega^1...$  be an infinite word
  over $2^{At}$. We write $\omega_i$ for the suffix of $\omega$ starting at $\omega^i$. We denote $\omega \models \varphi$ to express that
$\omega$ satisfies a formula $\varphi$:


\footnotesize{
 
$	\omega \models A \iff A \in \omega^0$

$ \omega \models \neg \varphi \iff \omega \nvDash \varphi $

$	\omega \models \varphi_1   {\vee} \varphi_2 \iff \omega \models \varphi_1 \text{ or } \omega\models \varphi_2$

$\omega\models X \varphi \iff \omega_1 \models \varphi$

$\omega \models \varphi_1 U \varphi_2 \iff \exists i \geq0, \omega_i \models\varphi_2 \text{ and } \forall 0 \leq j < i, \omega_j \models \varphi_1$
}
\normalsize

 The  temporal operators  G (globally) and F (eventually) are defined as follows: $   {F \varphi= (A \vee \neg A) U \varphi}$ and $G \varphi =\neg F \neg \varphi$. 
Let  $W(\varphi)$ be  the set of infinite words that satisfy  an LTL formula   $\varphi$.
It is well known that  $W(\varphi)$ can be accepted by B\"{u}chi automata:

\begin{definition}
	A B\"{u}chi automaton $\mathcal{B}$ is a quintuple $(Q,\Gamma,\eta,q_0,F)$ where $Q$ is  a finite set of states, $\Gamma$ is a finite input alphabet, $\eta \subseteq (Q \times \Gamma \times Q)$ is a set of transitions, $q_0 \in Q$ is the initial state and   {$F\subseteq Q$ is the set of accepting states. A run of $\mathcal{B}$ on a 
word $\gamma_0\gamma_1... \in \Gamma^{\omega}$ is a sequence of states  $q_0q_1q_2...$  s.t. $\forall i\geq 0, (q_i,\gamma_i,q_{i+1})\in \eta$.} An infinite word $\omega$  is accepted by $\mathcal{B}$  if $\mathcal{B}$ has a run on 
$\omega$ that starts at $q_0$ and visits accepting states from $F$ infinitely often. 
\end{definition}

\begin{citetheorem}\cite{LTL}
Given an LTL formula $\varphi$, one can effectively construct a B\"{u}chi automaton $\mathcal{B}_{\varphi}$ which accepts  $W(\varphi)$.
\end{citetheorem}

\subsection{Self Modifying B\"{u}chi Pushdown Systems} 
\begin{definition}
 		A Self Modifying B\"{u}chi Pushdown Systems (SM-BPDS) is a tuple $\mathcal{BP}=(P,\Gamma, \Delta,\Delta_c,G)$ where  { $P$ is a set of control locations, $G\subseteq P$ is a set of accepting control locations, $\Delta\subseteq  (P\times \Gamma) \times (P\times \Gamma^*)$ is a finite set of transition rules, and $\Delta_c \subseteq P\times 2^{\Delta \cup \Delta_c} \times 2^{\Delta \cup \Delta_c} \times P$ is a finite set of modifying transition rules in the form $p\smrule{(\sigma,\sigma')}p'$ where $\sigma, \sigma' \subseteq \Delta \cup \Delta_c$. }

Let $\Rightarrow_{\mathcal{BP}}$ be the  {transition relation between configurations as follows: Let $\theta\subseteq \Delta\cup \Delta_c,\gamma \in \Gamma, w\in \Gamma^*$, and $p\in P$, then}

\begin{enumerate}
	\item  {If $r: \langle p,\gamma \rangle \hookrightarrow \langle p',w' \rangle \in \Delta$ and $r\in \theta$, then $(\langle p,\gamma w \rangle, \theta) \Rightarrow_{\mathcal{BP}} (\langle p',w' w \rangle,\theta)$.}
	\item  {If $r: p\smrule{(\sigma,\sigma')} p' \in \Delta_c$,  {$\sigma \cap \theta \neq \emptyset$} and $r\in \theta$, then $(\langle p,\gamma w \rangle, \theta)\Rightarrow_{\mathcal{BP}} (\langle p',\gamma w \rangle, \theta')$ where $\theta'=\theta \backslash \sigma \cup \sigma'$.}
\end{enumerate}

 {A run $\pi$  of $\mathcal{BP}$ is a sequence of configurations $\pi=c_0c_1...$  s.t.  
$c_i \Rightarrow_{\mathcal{BP}} c_{i+1}$ for every $i\geq 0$. $\pi$  is accepting iff it infinitely often visits configurations having  control locations in $G$.}

 Let $c$ and $c'$ be two  configurations of the SM-BPDS $\mathcal{BP}$. The relation $\Rightarrow_{\mathcal{BP}}^r$  is defined as follows: $c \Rightarrow_{\mathcal{BP}}^r c'$ iff there exists a configuration $(\langle g, u\rangle,\theta)$, $g\in G$ s.t. 
$c\Rightarrow_{\mathcal{BP}}^* (\langle g, u\rangle,\theta) \Rightarrow_{\mathcal{BP}}^+ c'$. We remove the subscript $\mathcal{BP}$ when it is clear from the context. {We define $ \stackrel{i}{\Rightarrow}$ as follows: $c \stackrel{i}{\Rightarrow} c'$ iff there exists  a sequence of configurations $c_0\Rightarrow_{\mathcal{BP}} c_1 \Rightarrow_{\mathcal{BP}}...\Rightarrow_{\mathcal{BP}} c_i$ s.t. $c_0=c$ and $c_i=c'$.}

 	A head of SM-BPDS is a tuple $(\langle p, \gamma \rangle,\theta)$ where $p\in P$, $\gamma\in \Gamma$ and $\theta \subseteq \Delta \cup \Delta_c$. A head $(( p,\gamma ),\theta)$ is repeating if there exists $v \in \Gamma^*$ such that $(\langle p,\gamma \rangle, \theta) \Rightarrow_{\mathcal{BP}}^r (\langle p,\gamma v \rangle,\theta)$.
 The set of repeating heads of SM-BPDS is called  $Rep_{\mathcal{BP}}$.
 	\end{definition} 

 { We assume w.l.o.g. that for every rule in $\Delta_c$ of the form $r:p\smrule{(\sigma,\sigma')} p'$, $r \notin \sigma.$}
\subsection{From LTL Model-Checking of SM-PDSs to the emptiness problem of SM-BPDSs}
Let $\mathcal{P}=(P,\Gamma,\Delta,\Delta_c)$ be a self modifying pushdown system. 
Let $At$ be a set of atomic propositions. Let $\nu: P \rightarrow 2^{At}$ be a  labelling function. Let $\pi=(\langle p_0,w_0 \rangle,\theta_0)(\langle p_1,w_1 \rangle,\theta_1)...$ be an execution of the SM-PDS $\mathcal{P}$. Let $\varphi$ be an LTL formula over the set of atomic propositions $At$. We say that 
$$\pi\models_{\nu}\varphi \mbox{  iff  }  \nu(p_0) \nu(p_1)\cdots \models\varphi$$

Let $(\langle p, w \rangle, \theta)$ be a configuration of $\mathcal{P}$. We say that $(\langle p, w \rangle, \theta) \models_{\nu}\varphi$ iff $\mathcal{P}$
has a path $\pi$ starting at $(\langle p, w \rangle, \theta)$ such that $\pi\models_{\nu}\varphi$.

\medskip

Our goal in this paper is to perform LTL model-checking for self-modifying pushdown systems.
Since SM-PDSs can be translated to standard (symbolic) pushdown systems, one way to solve this LTL model-checking problem is to compute the (symbolic) pushdown system
that is  equivalent to the SM-PDS (see section \ref{versPDS}), and then apply the standard LTL model-checking  algorithms on standard PDSs  \cite{Schwoon:2007vs}.
However, this approach is not efficient (as will be witnessed later in the experiments). Thus, we need a {\em direct} approach that performs LTL model-checking on the 
SM-PDS, without translating it to an equivalent PDS.
Let   $\mathcal{B}_{\varphi} =(Q,2^{At}, \eta, q_0,F)$  be a B{\"u}chi automaton that accepts $W(\varphi)$.
We compute the SM-BPDS $\mathcal{BP}_{\varphi}=(P \times Q, \Gamma, \Delta',\Delta_c', G)$  by performing a kind of product between the SM-PDS $\mathcal{P}$ and 
the B\"uchi automaton $\mathcal{B}_{\varphi}$ as follows:

\begin{enumerate}
\item  if $r= \langle p,\gamma \rangle \hookrightarrow \langle p', w \rangle \in \Delta$ and $(q,   {\nu(p)}, q')\in \eta$, then  
$\langle(p,q),\gamma \rangle \hookrightarrow \langle(p',q'),w \rangle \in \Delta'$. Let $prod(r)$ be the set of rules of $\Delta'$  obtained from the rule $r$, 
i.e., rules of $\Delta'$ of the form $\langle(p,q),\gamma \rangle \hookrightarrow \langle(p',q'),w \rangle$.
\item if a rule $r= p \smrule{(r_1,r_2)}p'\in \Delta_c$ and $(q,   {\nu(p)}, q')\in \eta$, then   {$(p,q) \smrule{(\sigma,\sigma')} (p',q') \in \Delta_c'$ where}  {$\sigma = prod(r_1),\sigma'=prod(r_2)$}. 
Let $prod(r)$ be the set of rules of $\Delta'$  obtained from the rule $r$,  i.e., rules of $\Delta_c'$ of the form  {$(p,q) \smrule{(\sigma,\sigma')} (p',q')$}. 
\item $G=P\times F$.
\end{enumerate} 

We can show that:
\begin{theorem}
Let  $(\langle p,w\rangle,\theta)$ be a configuration of the SM-PDS $\mathcal{P}$.
$(\langle p,w\rangle,\theta) \models_{\nu} \varphi$ iff $\mathcal{BP}_{\varphi}$ has an accepting run from $(\langle(p,q_0),w\rangle,   {prod(\theta))}$ where $prod(\theta)$ is the set of rules of $\Delta \cup \Delta_c$ obtained from the rules of $\theta$ as described  above.
\end{theorem}

Thus, LTL model-checking for SM-PDSs can be reduced to checking whether a SM-BPDS has an accepting run.
The rest of the paper is devoted to this problem.

\section{The Emptiness Problem of SM-BPDSs} 

From now on, we fix a SM-BPDS $\mathcal{BP}=(P,\Gamma, \Delta,\Delta_c,G)$. 
We can show that  $\mathcal{BP}$ has an accepting run starting from a configuration $c$ if and only if from $c$, it can reach a configuration with a
repeating head:

  	 	\begin{proposition}
 	\label{prop}
A  SM-BPDS $\mathcal{BP}$ has an accepting run starting from a configuration $c$ if and only if there exists a repeating head $(( p,\gamma ), \theta)$ such that $c \Rightarrow_{\mathcal{BP}}^* (\langle p,\gamma w \rangle, \theta)$ for some $w \in \Gamma^*$. 
 	\end{proposition}
\begin{Proof}
 		$``\Rightarrow"$: Let $\sigma=c_0c_1...$ be an accepting run  starting at configuration $c$ where $c_0=c$ and $c_i=(\langle p_i,w_i \rangle,\theta_i)$. We construct an increasing sequence of indices $i_0,i_1...$ with a property that once any of the configurations $c_{i_k}$ is reached, the rest of the run never changes the bottom $\lvert w_{i_k} \lvert-1$ elements of the stack anymore. This property can be written as follows:
 		\begin{equation*}
 			\lvert w_{i_0} \lvert=\text{min}\{ \lvert w_j \lvert~|~j\geq 0 \}
 		\end{equation*} 
 		\begin{equation*}
 			 \lvert w_{i_k} \lvert=\text{min}\{ \lvert w_j \lvert~|~j> i_{k-1}\}, k\geq 1 
 		\end{equation*} 
 		Because  $\mathcal{BP}$ has only finitely many different heads, there must be a head $(\langle p,\gamma\rangle,\theta)$ which  occurs infinitely often as a head in the sequence $c_{i_0}c_{i_1}...$. Moreover, as some $g\in G$ becomes  a  control location infinitely often, we can find a subsequence of indices $i_{j_0},i_{j_1},...$ with the following property: for every $k\geq 1,$ there exist $v,w\in \Gamma^*$
 		\begin{equation*}
 			c_{i_{j_k}}=(\langle p,\gamma w\rangle,\theta) \Rightarrow^r (\langle p,\gamma v w\rangle,\theta)=c_{i_{j_{k+1}}} 
 		\end{equation*}
 		Because $w$ is never looked at or changed in this path, we can have $(\langle p,\gamma \rangle,\theta) \Rightarrow^r (\langle p,\gamma v \rangle,\theta)$.  This proves this direction of the proposition.
 		
 		\medskip
 		\noindent
 		$``\Leftarrow"$: Because $(\langle p,\gamma \rangle,\theta)$ is a repeating head, we can construct the following run  for some $u,v,w\in \Gamma^*,$ $\theta' \subseteq (\Delta \cup \Delta_c) $ and $g\in G$:
 		\small
 		\begin{equation*}
 			c\Rightarrow^*(\langle p,\gamma w \rangle,\theta)\Rightarrow^*(\langle g,uw \rangle,\theta')\Rightarrow^+(\langle p,\gamma v w \rangle,\theta) \Rightarrow^*(\langle g,u v w \rangle,\theta') \Rightarrow^+(\langle p,\gamma v v w \rangle,\theta) \Rightarrow^* ... 
 		\end{equation*}
 		\normalsize
 		Since $g$ occurs infinitely often, the run is accepting.
 	\end{Proof}

Thus, since there exists an efficient algorithm to compute the $pre^*$ of SM-PDSs \cite{touili2017reachability},
the emptiness problem of a SM-BPDS can be reduced to computing  its repeating heads.

\subsection{The Head Reachability Graph $\mathcal{G}$}

Our goal is to compute the set of repeating heads  $Rep_{\mathcal{BP}}$, i.e., the  set of heads $(\langle p,\gamma \rangle,\theta)$
such that there exists $v\in\Gamma^*$,   $(\langle p,\gamma \rangle,\theta) \Rightarrow^r (\langle p,\gamma v\rangle,\theta)$. I.e., 
 $(\langle p,\gamma \rangle,\theta) \Rightarrow^* (\langle p,\gamma v\rangle,\theta)$ s.t. this path goes through an accepting location in $G$.
To this aim, we will compute a finite graph $\mathcal{G}$ whose  nodes are the heads of $\mathcal{BP}$ of the form  $((p,\gamma),\theta)$, 
where $p\in P$, $\gamma\in \Gamma$ and $\theta\subseteq\Delta\cup\Delta_c$; and whose edges encode the reachability relation between these heads.
More precisely, given two heads $((p,\gamma),\theta)$ and $((p',\gamma' ),\theta')$, 
$((p,\gamma),\theta) \xrightarrow{b} ((p',\gamma' ),\theta')$  is an edge of the graph $\mathcal{G}$ 
means that the configuration  $(\langle p,\gamma \rangle,\theta)$ can reach a configuration having $(\langle p',\gamma' \rangle,\theta')$ as head, i.e., it means that 
 there exists 
$v\in\Gamma^*$ s.t.  $(\langle p,\gamma \rangle,\theta) \Rightarrow^* (\langle p',\gamma' v\rangle,\theta')$.
Moreover, we need to keep the information whether this path visits an accepting location in $G$ or not. This information is recorded in the label of the edge $b$:
$b=1$ means that  the path visits an accepting location in $G$, i.e. that 
$(\langle p,\gamma \rangle,\theta) \Rightarrow^r (\langle p',\gamma' v\rangle,\theta')$.
Otherwise, $b=0$.
Therefore, if the graph $\mathcal{G}$ contains a loop from a head $((p,\gamma),\theta)$ to itself such that this loop goes through an edge labelled by $1$,
then $((p,\gamma),\theta)$ is a  repeating head. Thus, computing  $Rep_{\mathcal{BP}}$ can be reduced to computing the graph $\mathcal{G}$ and finding 
1-labelled loops in this graph.

More precisely, we define the  head reachability graph $\mathcal{G}$ as follows:

\begin{definition}
\label{graph}

The  head reachability graph $\mathcal{G}$ is a tuple $(P \times \Gamma \times 2^{\Delta \cup \Delta_c},\{0,1\}, \delta)$ such that 
$((p,\gamma),\theta) \xrightarrow{b} ((p',\gamma'),\theta')$ is an  edge of $\delta$ iff:

\begin{enumerate}
\item  there exists a transition $r_c: p$  {$\smrule{(\sigma,\sigma')} $} $p' \in \theta\cap\Delta_c $,  $\gamma=\gamma'$, $ { \theta'=\theta\setminus \sigma \cup \sigma'}$, and 
$b=1$ iff $p \in G$; 

\item there exists a transition $\langle p,\gamma \rangle \hookrightarrow \langle p',\gamma'\rangle \in \theta \cap \Delta,\theta=\theta'$ and $b=1$ iff $p\in G$;

\item  there exists a transition $ \langle p, \gamma \rangle \hookrightarrow \langle p'', \gamma_1 \gamma'  \rangle\in \theta\cap\Delta$, for $\gamma_1\in\Gamma$, $p''\in P$, 
s.t. $(\langle p'',\gamma_1 \rangle,\theta) \Rightarrow_{\mathcal{BP}}^* (\langle p',\epsilon \rangle ,\theta')$, and $b=1$ iff $p \in G$ or 
$(\langle p'',\gamma_1 \rangle,\theta) \Rightarrow_{\mathcal{BP}}^r (\langle p',\epsilon \rangle ,\theta')$
\end{enumerate}
		{Let $\mathcal{G}$ be the head reachability graph. We define $\xrightarrow[i]{}$ as follows: let $((p,\gamma),\theta)$ and $((p',\gamma'),\theta')$ be two heads of $\mathcal{BP}$. We write $((p,\gamma),\theta)\xrightarrow[i]{}((p',\gamma'),\theta')$ iff $\exists $ booleans $b_1,b_2...b_i \in \{ 0,1\}$, $\exists $ heads $((p_j,\gamma_j),\theta_j), 0\leq j\leq i$ s.t.  $\mathcal{G}$ contains the following path $((p_0,\gamma_0),\theta_0) \xrightarrow{b_1}((p_1,\gamma_1),\theta_1) \xrightarrow{b_2}...\xrightarrow{b_i}((p_i,\gamma_i),\theta_i)$ where $((p_0,\gamma_0),\theta_0)=((p,\gamma),\theta)$ and $((p_i,\gamma_i),\theta_i)=((p',\gamma'),\theta')$.}
  		
  		\medskip
  		{Let $\rightarrow^*$ be the reflexive transitive closure of the graph relation $\xrightarrow{b}$, and let $\rightarrow^r$ be defined as follows: Given two heads
 $((p,\gamma),\theta)$ and  $((p',\gamma'),\theta')$,
$((p,\gamma),\theta) \rightarrow^r ((p',\gamma'),\theta')$  iff there is in $\mathcal{G}$ a path  between  $((p,\gamma),\theta)$ and  $((p',\gamma'),\theta')$ that 
goes through a 1-labelled edge, i.e., iff there exist heads $((p_1,\gamma_1),\theta_1)$ and $ ((p_2,\gamma_2),\theta_2) $ s.t. 
	$((p,\gamma),\theta) \rightarrow^* ((p_1,\gamma_1),\theta_1) \xrightarrow{1} ((p_2,\gamma_2),\theta_2) \rightarrow^* ((p',\gamma'),\theta').$}
\end{definition}

We can show that:

		\begin{theorem}
		\label{theorem41}
 		Let $\mathcal{BP}=(P,\Gamma,\Delta,\Delta_c,G)$  be  a self-modifying B\"{u}chi pushdown system, and let $\mathcal{G}$ be its corresponding head reachability graph. A head $(( p,\gamma),\theta)$ of $\mathcal{BP}$ is repeating iff $\mathcal{G}$ has a loop on the node $((p,\gamma),\theta)$  that goes through a  1-labeled edge.
 	\end{theorem} 

To prove this theorem, we first need to prove the following lemma:
 	\begin{lemma}
 	\label{lemma1}
 		The relations $\rightarrow^*$ and $\rightarrow^r$ have the following properties: For any heads $((p,\gamma),\theta_1)$ and $ ((p',\gamma'),\theta_2)$:
 		\begin{enumerate}
 			\item[(a)] $((p,\gamma),\theta_1) \rightarrow^* ((p',\gamma'),\theta_2)$ iff $(\langle p,\gamma\rangle,\theta_1) \Rightarrow^* (\langle p',\gamma' v\rangle,\theta_2)$ for some $v\in \Gamma^*$.
 			\item[(b)] $((p,\gamma),\theta_1) \rightarrow^r ((p',\gamma'),\theta_2)$  iff $(\langle p,\gamma\rangle,\theta_1) \Rightarrow^r (\langle p',\gamma' v\rangle,\theta_2)$ for some $v\in \Gamma^*$. 
 		\end{enumerate}
 	\end{lemma}
 	\begin{Proof}
 	``$\Rightarrow$": Assume $((p,\gamma),\theta_1) \xrightarrow[i]{} ((p',\gamma'),\theta_2)$. We proceed by induction on $i$. 
 		\begin{enumerate}
 			\item[(a)] \textbf{Basis.} $i=0$. In this case, $((p,\gamma),\theta_1) = ((p',\gamma'),\theta_2)$, then we can get $(\langle p,\gamma\rangle,\theta_1) \Rightarrow^* (\langle p,\gamma\rangle,\theta_1)=(\langle p',\gamma'\rangle,\theta_2)$
 
 \medskip
 			 \textbf{Step.} $i>0$. Then there exist $p_1\in P,\gamma'' \in \Gamma^*$ and $\theta'\subseteq \Delta \cup \Delta_c$ such that $((p,\gamma),\theta_1) \xrightarrow[1]{}((p_1,\gamma''),\theta') \xrightarrow[i-1]{}((p',\gamma'),\theta_2)$. From the induction hypothesis, there exists $u\in \Gamma^*$ such that  
 			$(\langle p_1,\gamma''\rangle,\theta') \Rightarrow^* (\langle p',\gamma' u\rangle,\theta_2)$
 			 			 
 			 Since $((p,\gamma),\theta_1) \rightarrow((p_1,\gamma''),\theta')$,  we have $	(\langle p,\gamma \rangle ,\theta_1) \Rightarrow^* ( \langle p_1,\gamma'' w\rangle, \theta')$ for $w \in \Gamma^*$, hence $(\langle p,\gamma \rangle ,\theta_1)  \Rightarrow^* (\langle p',\gamma' uw \rangle,\theta_2)$.

 			  The property holds. 
 			  \medskip \medskip
 			\item[(b)]  $((p,\gamma),\theta_1) \rightarrow^r ((p,\gamma),\theta_1)$ cannot hold for the case $i=0$.
 			
 			\medskip
 			\textbf{Basis.} $i=1.$ In this case,  $((p,\gamma),\theta_1) \rightarrow^r ((p',\gamma'),\theta_2)$, then we can get $p\in G$ and $(\langle p,\gamma\rangle,\theta_1) \Rightarrow^r (\langle p',\gamma'\rangle,\theta_2)$. The property holds.
 			 
 \medskip
 			 \textbf{Step.} $i>0$. As done in the proof of part (a) of this lemma, there exists $p_1,\gamma''\in \Gamma,\theta''\subseteq \Delta \cup \Delta_c$ s.t.   $((p,\gamma),\theta_1) \xrightarrow[1]{}((p_1,\gamma''),\theta') \xrightarrow[i-1]{}((p',\gamma'),\theta_2)$. Then if $((p,\gamma),\theta_1) \rightarrow^r ((p',\gamma'),\theta_2)$, either $((p_1,\gamma''),\theta') \rightarrow^r((p',\gamma'),\theta_2)$  or $((p,\gamma),\theta_1) \xrightarrow{1}((p_1,\gamma''),\theta')$ holds. 
 			  In the first case i.e. $((p_1,\gamma''),\theta') \rightarrow^r((p',\gamma'),\theta_2)$, by the induction hypothesis, we can have $(\langle p_1,\gamma''\rangle,\theta') \Rightarrow^r (\langle p',\gamma' u\rangle,\theta_2)$,
 			  hence, $(\langle p,\gamma\rangle,\theta_1) \Rightarrow^r (\langle p',\gamma' u\rangle,\theta_2)$ holds
\medskip
 			 
 			 The second case depends on the rule applied to get $((p,\gamma),\theta_1) \xrightarrow{1}((p_1,\gamma''),\theta')$ according to Definition \ref{graph}.
 			 
 			 \begin{itemize}
 			 	 			 	\item[-] If this edge corresponds to a transition $r_c:p$  {$\smrule{(\sigma,\sigma')}$} $p_1\in \theta_1$, then  $\gamma=\gamma'',\theta' =\theta_1 \backslash  {\sigma \cup \sigma'}$ and $p\in G$. Since we can obtain  $(\langle p,\gamma \rangle ,\theta_1) \Rightarrow_{\mathcal{BP}} (\langle p_1,\gamma \rangle ,\theta') \Rightarrow^* (\langle p',\gamma' uw \rangle,\theta_2)$ from part $(a)$ and $p\in G$, then $(\langle p,\gamma \rangle ,\theta_1) \Rightarrow^r (\langle p_1,\gamma \rangle ,\theta') \Rightarrow^* (\langle p',\gamma' uw \rangle,\theta_2)$. This implies that $(\langle p,\gamma \rangle ,\theta_1) \Rightarrow^r  (\langle p',\gamma' v \rangle,\theta_2)$ for some $v\in \Gamma^*.$
 	
 	\medskip
 			 	
 			 	\item[-] If this edge corresponds to a transition $r:\langle p,\gamma \rangle \hookrightarrow \langle p_1,\gamma'' \rangle \in \theta_1 \cap \Delta$, then $\theta'=\theta_1$ and $p\in G$.  Since we can obtain  $(\langle p,\gamma \rangle ,\theta_1) \Rightarrow_{\mathcal{BP}} (\langle p_1,\gamma'' \rangle,\theta_1) \Rightarrow^* (\langle p',\gamma' uw \rangle,\theta_2)$ from part $(a)$ and $p\in G$,  then $(\langle p,\gamma \rangle ,\theta_1) \Rightarrow^r (\langle p_1,\gamma'' \rangle,\theta_1) \Rightarrow^* (\langle p',\gamma' uw \rangle,\theta_2)$. This implies that   $(\langle p,\gamma \rangle ,\theta_1) \Rightarrow^r (\langle p',\gamma' v \rangle,\theta_2)$ for some $v\in \Gamma^*$.

\medskip
 			 	
 			 	\item[-]  {If this edge corresponds to a transition $r:\langle p,\gamma\rangle \hookrightarrow \langle p'', \gamma_1 \gamma'' \rangle\in \theta_1$, then either $p\in G$ or $(\langle p'',\gamma_1 \rangle,\theta_1) \Rightarrow^r (\langle p_1,\epsilon \rangle,\theta')$ holds. If $p\in G$, then we have $	(\langle p,\gamma \rangle ,\theta_1) \Rightarrow^r ( \langle p'',\gamma_1 \gamma''  \rangle, \theta_1)$. Otherwise, $( \langle p'',v_1 \gamma'' w \rangle, \theta_1) \Rightarrow^r (\langle p_1, \gamma'' w\rangle, \theta')$.  Since we can obtain $(\langle p_1,\gamma''\rangle,\theta') \Rightarrow^* (\langle p',\gamma' u\rangle,\theta_2)$ from part $(a)$. Therefore,  $(\langle p,\gamma \rangle ,\theta_1) \Rightarrow^r( \langle p_1,\gamma'' \rangle, \theta') \Rightarrow^* (\langle p',\gamma' u  \rangle,\theta_2)$. This implies that  $(\langle p,\gamma \rangle ,\theta_1) \Rightarrow^r(\langle p',\gamma' v  \rangle,\theta_2)$ for some $v\in \Gamma^*$.}

 			 \end{itemize}

 		\end{enumerate}
 		
 		 \medskip
 	
 	\noindent	 
 	`$\Leftarrow$": Assume $(\langle p,\gamma\rangle,\theta_1) \stackrel{i}{\Rightarrow} (\langle p',\gamma'v\rangle,\theta_2)$. We proceed by induction on $i$. 
 		\begin{enumerate}
 			\item[(a)] \textbf{Basis.} $i=0$. 
 	In this case, $v=\epsilon$ and $(\langle p,\gamma\rangle ,\theta_1)=(\langle p',\gamma'\rangle,\theta_2)$, then $((p,\gamma),\theta_1) \rightarrow^*((p',\gamma'),\theta_2)$ holds.
   \medskip
 			
 			 \textbf{Step.} $i>0$. Then there exist $p_1\in P,u\in \Gamma^*$ and $\theta'\subseteq \Delta\cup\Delta_c$ such that $(\langle p,\gamma\rangle,\theta_1) \stackrel{1}{\Rightarrow} (\langle p_1,u\rangle,\theta')\stackrel{i-1}{\Rightarrow} (\langle p',\gamma' v\rangle,\theta_2)$. There are 2 cases:
 	\medskip
 	\begin{enumerate}
 	 \item[1.] Case $\theta'=\theta_1:$ There must exist a rule $r: \langle p,\gamma \rangle \hookrightarrow \langle  p_1,u\rangle\in \Delta$ such that $r\in \theta'$ and  $|u|\geq 1$. Let $l$ denote the minimal length of the stack on the path from $(\langle p_1,u \rangle, \theta_1)$ to $(\langle p',\gamma' v\rangle,\theta_2)$. Then $u$ can be written as $u''\gamma_1 u'$ where $|u'|=l-1$ (that means $u'$ will remain on the stack for the path). Furthermore, there exists $p'''$ such that $(\langle p_1,u''\rangle,\theta_1) \Rightarrow^*(\langle p''',\epsilon \rangle,\theta'')$ for some $\theta'' \subseteq (\Delta_c \cup \Delta)$.  {We have $(\langle p,\gamma\rangle,\theta_1)\stackrel{k}{\Rightarrow}(\langle p''',\gamma_1 u' \rangle,\theta'')$ for $k<i$. By the induction on $i$, we have $((p,\gamma),\theta_1) \rightarrow^* ((p''',\gamma_1),\theta'')$.} Because $u'$ has to remain on the stack for the rest of the path, $v$ is of the form $v'u'$ for some $v'\in \Gamma^*$. That means $(\langle p''',\gamma_1\rangle, \theta'')\stackrel{j}{\Rightarrow} (\langle  p',\gamma' v'\rangle, \theta_2)$ for $j<i$. By the induction hypothesis, $((p''',\gamma_1),\theta'')\rightarrow^* ((p',\gamma'),\theta_2)$ holds. Moreover, we have $((p,\gamma),\theta_1) \rightarrow^* ((p''',\gamma_1),\theta'')$, hence $((p,\gamma),\theta_1) \rightarrow^*((p',\gamma'),\theta_2)$.
 	 \medskip
 	 \item[2.] Case $\theta'\neq\theta_1:$ There must be a rule $r_c: p$  {$ \smrule{(\sigma,\sigma')}$ } $p_1\in \Delta_c$ such that $r_c \in \theta_1$ and $ {\sigma \cap \theta_1 \neq \emptyset}$, then $\theta'= \theta_1 \setminus  {\sigma \cup \sigma'}$.  After the execution of $r_c$, the content of the stack will remain the same, thus, $u=\gamma$. Then  $(\langle p,\gamma\rangle,\theta_1) \stackrel{1}{\Rightarrow} (\langle p_1,\gamma\rangle,\theta')\stackrel{i-1}{\Rightarrow} (\langle p',\gamma' v\rangle,\theta_2)$. By the induction hypothesis to $(\langle p_1,\gamma\rangle,\theta')\stackrel{i-1}{\Rightarrow} (\langle p',\gamma' v\rangle,\theta_2)$, we can obtain that $((p_1,\gamma),\theta')\rightarrow^* ((p',\gamma'),\theta_2)$. Since $(\langle p,\gamma\rangle,\theta_1) \stackrel{1}{\Rightarrow} (\langle p_1,\gamma\rangle,\theta')$, then we can have a path $(( p,\gamma),\theta_1) \rightarrow (( p_1,\gamma),\theta')\rightarrow^*((p',\gamma'),\theta_2)$ that implies $(( p,\gamma),\theta_1) \rightarrow^*((p',\gamma'),\theta_2)$. The property holds.  	\end{enumerate}
 	
 			  \medskip \medskip
 			\item[(b)] $(\langle p,\gamma\rangle,\theta_1) \Rightarrow^r (\langle p,\gamma' v\rangle,\theta_1)$ is impossible in $0$ steps.
 	
 	\medskip
 	 \textbf{Basis.} $i=1$. $(\langle p,\gamma\rangle,\theta_1) \Rightarrow^r (\langle p,\gamma \rangle,\theta_1)$, then $p\in G$. Thus, $((p,\gamma),\theta_1) \rightarrow^r ((p,\gamma),\theta_1)$ holds.
 
       \medskip
 			 \textbf{Step.} $i>1$.  {$(\langle p,\gamma\rangle,\theta_1) \Rightarrow^r (\langle p',\gamma' v\rangle,\theta_2)$ holds,   then there exist $p_1\in P,u\in \Gamma^*$ and $\theta'\subseteq \Delta\cup\Delta_c$ such that $(\langle p,\gamma\rangle,\theta_1) \stackrel{1}{\Rightarrow} (\langle p_1,u\rangle,\theta')\stackrel{i-1}{\Rightarrow} (\langle p',\gamma' v\rangle,\theta_2)$. Thus, either $(\langle p,\gamma\rangle,\theta_1) \Rightarrow^r (\langle p_1,u\rangle,\theta')$ or $(\langle p_1,u\rangle,\theta') \Rightarrow^r (\langle p',\gamma'v \rangle,\theta_2)$ holds.} 
 	\medskip
 	
 	 The first case implies $p\in G.$ There are 2 cases:
 	 \begin{enumerate}
 	 	\item[1.] Case $\theta'=\theta_1:$ then as in the previous proof of part (a), we can have a path $((p,\gamma),\theta_1) \rightarrow^* ((p''',\gamma_1),\theta'') \rightarrow^* ((p',\gamma'),\theta_2)$.  { Since } $p\in G$, we get  {by Definition \ref{graph} $((p,\gamma),\theta_1)\rightarrow^*  ((p''',\gamma_1),\theta'')\rightarrow^*((p',\gamma'),\theta_2)$}.  Thus, we have that $((p,\gamma),\theta_1)\rightarrow^r ((p',\gamma'),\theta_2)$.  The property holds.
 	 	\item[2.] Case $\theta' \neq \theta_1$: then  as in the previous proof of part (a), we can have a path $(( p,\gamma),\theta_1) \rightarrow (( p_1,\gamma),\theta')\rightarrow^*((p',\gamma'),\theta_2)$. Since $p\in G$, we get $((p,\gamma),\theta_1)\xrightarrow{1}  (( p_1,\gamma),\theta')\rightarrow^*((p',\gamma'),\theta_2)$.  Thus, we have that $((p,\gamma),\theta_1)\rightarrow^r ((p',\gamma'),\theta_2)$.  The property holds.   
 	 \end{enumerate}

 			 \medskip
 			 { In the second case, $
 			 	(\langle p_1,u \rangle,\theta') \Rightarrow^r (\langle p',\gamma' v\rangle,\theta_2)$ holds. As previously, there are 2 cases:			 
 	   			\begin{enumerate}
 	   			\item[1.] Case $\theta'=\theta_1:$ then as in case (a) we have $(\langle p_1, u \rangle,\theta_1) \Rightarrow^* ( \langle p''', \gamma_1 u' \rangle,\theta'' )$   and $(\langle p''',\gamma_1\rangle, \theta'')\Rightarrow^* (\langle  p',\gamma' v'\rangle, \theta_2)$. If   $
 			 	(\langle p_1,u \rangle,\theta_1 ) \Rightarrow^r (\langle p',\gamma' v\rangle,\theta_2)$, then either $(\langle p_1,u\rangle,\theta_1)\Rightarrow^r (\langle p''',\gamma_1 u'\rangle,\theta'')$ or  $(\langle p''',\gamma_1  \rangle,\theta'') \Rightarrow^r (\langle p',\gamma' v' \rangle,\theta_2)$.
 	   			\begin{itemize}
 	   				\item[-]  If  $(\langle p_1,u\rangle,\theta_1)\Rightarrow^r (\langle p''',\gamma_1 u'\rangle,\theta'')$, let $u''\in \Gamma^*$ s.t. $u=u''\gamma_1u'$ and $(\langle p_1,u''\rangle,\theta_1)\Rightarrow^r (\langle p''',\epsilon \rangle,\theta'')$, then, we have  $((p,\gamma),\theta_1) \rightarrow^r ((p''',\gamma_1),\theta'')$. We have $(\langle p,\gamma\rangle,\theta_1)\stackrel{k}{\Rightarrow}(\langle p''',\gamma_1 u' \rangle,\theta'')$ for $k<i$. By the induction on $i$, we have $((p,\gamma),\theta_1) \rightarrow^* ((p''',\gamma_1),\theta'')$. Because $u'$ has to remain on the stack for the rest of the path, $v$ is of the form $v'u'$ for some $v'\in \Gamma^*$. That means $(\langle p''',\gamma_1\rangle, \theta'')\stackrel{j}{\Rightarrow} (\langle  p',\gamma' v'\rangle, \theta_2)$ for $j<i$. By the induction hypothesis, $((p''',\gamma_1),\theta'')\rightarrow^* ((p',\gamma'),\theta_2)$ holds. Moreover, we have $((p,\gamma),\theta_1) \rightarrow^* ((p''',\gamma_1),\theta'')$, hence $((p,\gamma),\theta_1) \rightarrow^*((p',\gamma'),\theta_2)$. So we can have a path $((p,\gamma),\theta_1)\rightarrow^*  ((p''',\gamma_1),\theta'')\rightarrow^*((p',\gamma'),\theta_2)$, thus we have that $((p,\gamma),\theta_1)\rightarrow^r ((p',\gamma'),\theta_2)$;
 	   				\item[-]If $(\langle p''',\gamma_1  \rangle,\theta'') \Rightarrow^r (\langle p',\gamma' v'\rangle,\theta_2)$, then by the induction hypothesis we have $((p''',\gamma_1),\theta'')\rightarrow^r ((p',\gamma'),\theta_2)$. Thus, we can have a path $((p,\gamma),\theta_1)\rightarrow^*  ((p''',\gamma_1),\theta'')\rightarrow^*((p',\gamma'),\theta_2)$, then we have that $((p,\gamma),\theta_1)\rightarrow^r ((p',\gamma'),\theta_2)$;
 	   			\end{itemize}
 	   			\item[2.] Case $\theta'\neq \theta_1:$ then  $(\langle p_1,\gamma\rangle,\theta') \Rightarrow^r (\langle p',\gamma' v\rangle,\theta_2)$. By the induction hypothesis we have $((p
 	   			_1,\gamma),\theta')\rightarrow^r ((p',\gamma'),\theta_2)$. Since $(\langle p,\gamma\rangle,\theta_1) \stackrel{1}{\Rightarrow} (\langle p_1,\gamma\rangle,\theta')\stackrel{i-1}{\Rightarrow} (\langle p',\gamma' v\rangle,\theta_2)$. 
 	   			
 	   			\noindent
 	   			By the induction hypothesis to $(\langle p_1,\gamma\rangle,\theta')\stackrel{i-1}{\Rightarrow} (\langle p',\gamma' v\rangle,\theta_2)$, we can obtain that $((p_1,\gamma),\theta')\rightarrow^* ((p',\gamma'),\theta_2)$. Since $(\langle p,\gamma\rangle,\theta_1) \stackrel{1}{\Rightarrow} (\langle p_1,\gamma\rangle,\theta')$, then we can have a path $(( p,\gamma),\theta_1) \rightarrow (( p_1,\gamma),\theta')\rightarrow^*((p',\gamma'),\theta_2)$. Thus, we have that $((p,\gamma),\theta_1)\rightarrow^r ((p',\gamma'),\theta_2)$;
 	   			\end{enumerate}}
 	   		
  			 Thus,   the property holds.
 		
 		\end{enumerate}
 	\end{Proof}
\textbf{Proof of Theorem \ref{theorem41}}
 	
 	\noindent
 	We can now prove Theorem \ref{theorem41}. 
 	
 	\noindent
 	\begin{Proof}
 	Let $(( p,\gamma),\theta)$ be a repeating head, then there exists some $v\in \Gamma^*, \theta \subseteq \Delta_c \cup \Delta$ such that $(\langle p,\gamma \rangle, \theta) \Rightarrow^r (\langle p,\gamma v\rangle,\theta)$.  {By Lemma \ref{lemma1}}, this is the case if and only if $((p,\gamma),\theta) \rightarrow^r((p,\gamma),\theta)$. From the definition of $\rightarrow^r$, that means that there exist heads $((p_1,\gamma_1),\theta') $  and $ ((p_2,\gamma_2),\theta'')$ such that $((p,\gamma),\theta) \rightarrow^* ((p_1,\gamma_1),\theta') \xrightarrow{1} ((p_2,\gamma_2),\theta'') \rightarrow^* ((p,\gamma),\theta).
$ Then $((p,\gamma),\theta), ((p_1,\gamma_1),\theta')$ and $((p_2,\gamma_2),\theta'')$ are all in  the same loop with a 1-labelled edge. Conversely, whenever $((p,\gamma),\theta)$ is in a component with such an edge, $((p,\gamma),\theta)\rightarrow^r ((p,\gamma),\theta)$ holds, then  { Lemma \ref{lemma1} implies that }$(\langle p,\gamma \rangle, \theta) \Rightarrow^r (\langle p,\gamma v\rangle,\theta)$ which means that $(( p,\gamma),\theta)$ is a repeating head.
	
 \end{Proof}

\subsection{Labelled configurations and labelled $\mathcal{BP}$-automata}

To compute  $\mathcal{G}$, we need to be able to compute  predecessors of configurations of the form $(\langle p',\epsilon \rangle ,\theta')$, and to 
determine whether these predecessors were backward-reachable using some control points in $G$ (item 3 in Definition \ref{graph}). 
To solve this question, we  will label configurations $(\langle p'',w \rangle,\theta)$ s.t. 
$(\langle p'',w \rangle,\theta) \Rightarrow^* (\langle p',\epsilon \rangle ,\theta')$ by  $1$ if  this path went through an accepting location in $G$, i.e.,
if  $(\langle p'',w \rangle,\theta) \Rightarrow^r (\langle p',\epsilon \rangle ,\theta')$, and by $0$ if not. To this aim, we define a labelled configuration
as a tuple $[(\langle p, w \rangle, \theta),b]$, s.t. $(\langle p, w \rangle, \theta)$ is a configuration and $b\in \{ 0,1\}$.

Multi-automata were introduced in \cite{Bouajjani:1997ew,esparza} to finitely represent regular infinite sets of configurations of a PDS.
Since a labelled configuration $c=[(\langle p, w \rangle, \theta),b]$ of a SM-PDS  involves a PDS configuration $\langle p,w\rangle$,
together with the current set of  transition rules (phase) $\theta$,  and a  boolean $b$, in order to take into account the phases $\theta$, and
 these new $0/1$-labels  in configurations,   we extend multi-automata to  labelled   {$\mathcal{BP}$}-automata as follows:

 \begin{definition}
 \label{labelledauto}
 	 Let $\mathcal{BP}=(P,\Gamma,\Delta,\Delta_c, G)$ be a SM-BPDS. A labelled  $  {\mathcal{BP}}$-automaton is a tuple  
$\mathcal{A}=(Q,\Gamma,T,I,F)$ where  $\Gamma$ is the automaton alphabet, $Q$ is a finite set of states, $I\subseteq P \times 2^{\Delta \cup \Delta_c}\subseteq Q$ is the set of initial states,   $T \subset  Q  \times \big((\Gamma\cup\{\epsilon\})\times \{ 0,1\} \big)\times  Q$ is the set of transitions, $F\subseteq Q$ is the set of final states.
 \end{definition}

\noindent
If $\big(q,[\gamma,b],q'\big)\in T$, we write $q \by{[\gamma,b]}_T q'$.
We extend this notation in the obvious way to sequences of symbols:
(1) $\forall q \in Q ,  q \by{[\epsilon,0]}_T q$, and (2)  $\forall q, q' \in Q ,   \forall b\in \{ 0,1\}, 
\forall w \in \Gamma^* \text{ for } w=\gamma_0...\gamma_{n+1},  q \by{[w,b]}_T q'$ iff  $\exists q_0,...,q_n \in Q, b_0,...,b_{n+1} \in \{ 0,1\},b=b_0\vee b_1\vee ...\vee b_{n+1}   \; \mbox{and}  \; q{\by{[\gamma_0,b_0]}}_T q_0{\by{[\gamma_1,b_1]}}_T q_1\cdots q_n{\by{[\gamma_{n+1},b_{n+1}]}}_T q'$. 
  { If $q\xrightarrow{[w,b]}_{T} q'$ holds, we say that $q\xrightarrow{[w,b]}_{T} q'$   { and $q{\by{[\gamma_0,b_0]}}_T q_0{\by{[\gamma_1,b_1]}}_T q_1\cdots q_n{\by{[\gamma_{n+1},b_{n+1}]}}_T q'$ }is a path of $\mathcal{A}$}.

A labelled configuration $[(\langle p, w \rangle, \theta),b]$ is  accepted by the automaton $\mathcal{A}$ iff there exists  a path
$(p,\theta){\by{[\gamma_0,b_0]}}_T q_1{\by{[\gamma_1,b_1]}}_T q_2\cdots q_n{\by{[\gamma_n,b_n]}}_T q_{n+1}$ in $\mathcal{A}$  such that 
$w=\gamma_0\gamma_1\cdots\gamma_n$, $b=b_0 \vee b_1\vee...\vee b_n$, $(p,\theta)\in I$, and $q_{n+1}\in F$. 
Let $L(\mathcal{A})$ be the set of labelled configurations accepted by $\mathcal{A}$.

\subsection{Computing $pre^*\big((\langle p',\epsilon \rangle ,\theta')\big)$}

Given a  configuration  
of the form   $(\langle p',\epsilon \rangle ,\theta')$,
our goal is to compute a labelled $\mathcal{BP}$-automaton 
$\mathcal{A}_{pre^*}\big((\langle p',\epsilon \rangle ,\theta')\big)$ that accepts labelled configurations
of the form $[c,b]$ where  $c$ is a configuration and $ b\in\{ 0,1\}$ such that $c\Rightarrow^*(\langle p',\epsilon \rangle ,\theta')$ 
(i.e., $c\in pre^*\big((\langle p',\epsilon \rangle ,\theta')\big)$) and 
$b=1$ iff  this path went through final control points, i.e., $c\Rightarrow^r(\langle p',\epsilon \rangle ,\theta')$. Otherwise, $b=0$.

\noindent
Let $p\in P$,  we define $B(p)=1$ if $p\in G$ and $B(p)=0$ otherwise.
$\mathcal{A}_{pre^*}\big((\langle p',\epsilon \rangle ,\theta')\big)=(Q,\Gamma,T, I, F)$ is computed as follows: 
Initially, $Q=I=F=\{(p',\theta')\}$ and  $T=\emptyset$. 
We add to $T$ transitions as follows:

\begin{enumerate}
\item[$\alpha_1$:] If $r=\langle p,\gamma \rangle \hookrightarrow \langle p_1,w \rangle \in \Delta$.
If  there exists   { in $T$} a path   {$(p_1,\theta)\by{[w,b]}_{T}q$ }  (in case $|w|=0$,
we have $w=\epsilon$) with $r\in \theta$. Then, add  $(p,\theta)$ to $I$, and $\big((p,\theta),[\gamma, B(p)\vee b],q \big)$ to $T$.

\item[$\alpha_2$:]  if  {$r=p \smrule{(\sigma,\sigma')} p_1 \in \Delta_c$} and there exists   {in $T$ }a transition   $(p_1,\theta)  {\by{[\gamma,b]}_{T}}q$ with $r\in \theta$, where
$\gamma\in\Gamma$. Then add  $(p,\theta')$ to $I$, and $\big((p,\theta'),[\gamma, B(p)\vee b],q\big)$ to $T$, for $\theta'$ such that 
 $ \theta=\theta'\setminus  {\sigma \cup \sigma'}$.    

\end{enumerate}

The  procedure above  terminates since there is a finite number of states and phases.
Note that  by construction, $F=\{(p',\theta') \}$, and,  since  initially $Q=\{(p',\theta') \}$, 
states of $\mathcal{A}_{pre^*}\big( (\langle p',\epsilon\rangle,\theta')\big)$ are all of the form $(p,\theta)$ for $p\in P$ and $\theta \subseteq \Delta \cup \Delta_c$.

\noindent
Let us explain the intuition behind  rule ($\alpha_1$).
 Let $r=\langle p,\gamma \rangle \hookrightarrow \langle p_1,w \rangle \in \Delta$. Let $c=(\langle p_1,ww'\rangle,\theta)$ and $c'=(\langle p,\gamma w'\rangle,\theta)$. Then, if $c\Rightarrow^* (\langle p',\epsilon \rangle,\theta')$, then necessarily, $c'\Rightarrow^* (\langle p',\epsilon \rangle,\theta')$. Moreover, $c'\Rightarrow^r (\langle p',\epsilon \rangle,\theta')$ iff either $c\Rightarrow^r (\langle p',\epsilon \rangle,\theta')$ or $p\in G$ (i.e. $B(p)=1$). Thus, we would like that if 
the automaton $\mathcal{A}_{pre^*}\big((\langle p',\epsilon \rangle,\theta')\big)$ accepts the labelled configuration $[c,b]$
(where $b=1$ means $c\Rightarrow^r (\langle p',\epsilon\rangle,\theta')$), then it should also accept the  labelled configuration $[c',b\vee B(p)]$
($b\vee B(p)=1$ means $c'\Rightarrow^r (\langle p',\epsilon\rangle,\theta')$). Thus, if 
the automaton $\mathcal{A}_{pre^*}\big((\langle p',\epsilon \rangle,\theta')\big)$
contains a path of the form $\pi =(p_1,\theta)\xrightarrow{[w,b_1]}_{T} q\xrightarrow{[w',b_2]}_{T} q_f$ where $q_f\in F$ that accepts the labelled configuration $[c,b]$,
then  the automaton should also accept the labelled configuration 
$[c',b\vee B(p)]$.
This configuration is accepted by the run  $(p,\theta)\by{[\gamma,B(p)\vee b_1]}_{T}q\by{[w',b_2]}_{T}q_f$ added by rule ($\alpha_1$).


Rule ($\alpha_2$) deals with modifying rules: Let  $r=p\smrule{(r_1,r_2)} p_1 \in \Delta_c$.  Let $c=(\langle p_1,\gamma w'\rangle,\theta)$ and $c'=(\langle p,\gamma w'\rangle,\theta'')$ s.t. $\theta=\theta''\backslash \{ r_1\} \cup \{r_2\}.$ Then, if $c\Rightarrow^* (\langle p',\epsilon \rangle,\theta')$, then necessarily, $c'\Rightarrow^* (\langle p',\epsilon \rangle, \theta')$. Moreover, $c'\Rightarrow^r (\langle p',\epsilon \rangle,\theta')$ iff either $c\Rightarrow^r (\langle p',\epsilon \rangle,\theta')$ or $p\in G$ (i.e. $B(p)=1$). Thus, we need to impose that if the automaton $\mathcal{A}_{pre^*}\big((\langle p',\epsilon \rangle,\theta')\big)$ contains a path of the form $(p_1,\theta) \xrightarrow{[\gamma,b_1]}_{T} q \xrightarrow{[w',b_2]}_{T} q_f$ (where $q_f\in F$) that accepts the labelled configuration $[c,b],b=b_1\vee b_2$ ($b=1$ means $c\Rightarrow^r (\langle p',\epsilon\rangle,\theta')$), then necessarily, the automaton $\mathcal{A}_{pre^*}\big( (\langle p',\epsilon \rangle, \theta') \big)$ should also accept the labelled configuration $[c',b\vee B(p)]$. This configuration is accepted by the run $(p,\theta'')\xrightarrow{[\gamma,B(p)\vee b_1]}_{T} q \xrightarrow{[w',b_2]}_{T} q_f$ added by rule ($\alpha_2$).

\bigskip

Before proving that our construction is correct, we introduce the following definition:
\begin{definition}
	Let $\mathcal{A}_{pre^*}\big( (\langle p',\epsilon\rangle,\theta')\big)=(Q,\Gamma,T,P,F)$ be the labelled $\mathcal{P}$-automaton computed by the saturation procedure above. In this section, we use $\xrightarrow[i]{}_{T}$ to denote the transition relation of $\mathcal{A}_{pre^*}\big( (\langle p',\epsilon\rangle,\theta')\big)$ obtained after adding $i$ transitions using the saturation procedure above. Let us notice that due to the fact that initially $Q=\{(p',\theta') \}$ and due to rules $(\alpha_1)$ and $(\alpha_2)$ that at step $i$ add only transitions of the form $(p,\theta)\xrightarrow{\gamma}_{T}q$ for a state $q$ that is already in the automaton at step $i-1$, then, states of $\mathcal{A}_{pre^*}\big( (\langle p',\epsilon\rangle,\theta')\big)$ are all of the form $(p,\theta)$ for $p\in P$ and $\theta \subseteq \Delta \cup \Delta_c$.
\end{definition}

We can  show that:
\begin{lemma}
\label{prelemma1}
Let $p,p'' \in P$ and $\theta,\theta''\subseteq \Delta \cup \Delta_c$. Let $w\in \Gamma^*$ and $b\in \{0,1\}$.   {If a path $(p,\theta)\by{[w,b]}_{T}(p'',\theta'')$ is in $\mathcal{A}_{pre^*}\big((\langle p',\epsilon \rangle ,\theta')\big)$,   then $(\langle p,w \rangle, \theta) \Rightarrow^* (\langle p'', \epsilon \rangle,\theta'')$}. Moreover, if $b=1$, then  $(\langle p,w \rangle,\theta) \Rightarrow^r (\langle p'',\epsilon \rangle,\theta'')$.

\end{lemma}
 \begin{Proof}
	 {Initially, the automaton contains no transitions.}  
Let $i$ be an index such that $(p,\theta)\xlongrightarrow[i]{[w,b]}_{T}(p'',\theta'')$ holds.  We proceed  by induction on $i$. 

 \textbf{Basis.} \textbf{$i=0$},  { then $(p'',\theta'')\xlongrightarrow[0]{[\epsilon,0]}_{T} (p'',\theta'')$. This means $p''=p'$, $\theta''=\theta'.$ Since initially $Q=\{(p',\theta')\}$, then $(\langle p'',\epsilon \rangle,\theta'')\Rightarrow^* (\langle p'',\epsilon \rangle,\theta'')$ always holds.}
	
	 \textbf{Step.} $i>0$. Let $t=\big((p_1,\theta_1), [\gamma,b_1], (p_0,\theta_0)\big)$ be the $i$-th transition added to $\mathcal{A}_{pre^*}$ and $j$ be the number of times that $t$ is used in the path  { $ (p,\theta){\xlongrightarrow[i]{[w,b]}}_{T}(p'',\theta'')$.} The proof is by induction on $j$. If $j=0$, then we have  {$(p,\theta){\xlongrightarrow[i-1]{[w,b]}}_{T}(p'',\theta'')$} in the automaton, and we apply the induction hypothesis (induction on $i$) then  { we obtain  $(\langle p,w\rangle, \theta) \Rightarrow^* (\langle p'',\epsilon \rangle,\theta'')$.}   So assume that $j>0$. Then, there exist $u,v\in \Gamma^*$, $b',b'' \in \{ 0,1\}$ such that $w=u\gamma v$, $b=b'\vee b_1 \vee b''$ and 

	\begin{equation*}
\tag{1}
(p,\theta) {\xlongrightarrow[i-1]{[u,b']}}_{T} (p_1,\theta_1) {\xlongrightarrow[i]{[\gamma,b_1]}}_{T}  {(p_0,\theta_0) {\xlongrightarrow[i]{[v,b'']}}_{T}(p'',\theta'')}
\end{equation*}

The application of the induction hypothesis (induction on $i$) to $(p,\theta) \xlongrightarrow[i-1]{[u,b']}_{T} (p_1,\theta_1)$ gives that 
\begin{equation*}
\tag{2}
(\langle p,u\rangle,\theta) \Rightarrow^* (\langle p_1, \epsilon\rangle,\theta_1), \text{ moreover, if }b'=1,(\langle p,u\rangle,\theta) \Rightarrow^r (\langle p_1, \epsilon\rangle,\theta_1)
\end{equation*}
There are 2 cases depending on whether transition $t$ was added by saturation rule $\alpha_1$ or $\alpha_2$.
\medskip

\begin{enumerate}
	\item Case $t$ was added  by rule $\alpha_1$: There exist $p_2\in P$ and $w_2\in \Gamma^*$ such that
	\begin{equation*}
	\tag{3}
	r=\langle p_1,\gamma \rangle \hookrightarrow \langle p_2, w_2 \rangle \in \Delta \cap \theta_1
	\end{equation*} 
	and $\mathcal{A}_{pre^*}$ contains the following path:
	\begin{equation*}
	\tag{4}
	\pi'=(p_2,\theta_1)\xlongrightarrow[i-1]{[w_2,b_2]}_{T}(p_0,\theta_0)\xlongrightarrow[i]{[v,b'']}_{T}(p'',\theta''),~~ b_1=b_2 \vee B(p_1)
	\end{equation*}

	Applying the transition rule $r$, we get that
	\begin{equation*}
		\tag{5}
		(\langle p_1, \gamma v\rangle,\theta_1) \Rightarrow (\langle p_2, w_2 v\rangle,\theta_1)
	\end{equation*}
	
	By induction on $j$ (since transition $t$ is used $j-1$ times in $\pi'$), we get from (4) that 
	\begin{equation*}
		\tag{6}
		(\langle p_2,w_2 v\rangle,\theta_1) \Rightarrow^* (\langle p'',\epsilon\rangle,\theta'')\text{ moreover, if } b_2\vee b''=1,(\langle p_2,w_2 v\rangle,\theta_1) \Rightarrow^r 	(\langle p'',\epsilon\rangle,\theta'')
			\end{equation*}

	Putting (2), (5) and (6) together, we can obtain that 
	
	\begin{equation*}
	(\langle p,w\rangle,\theta)=(\langle p, u \gamma v\rangle,\theta) \Rightarrow^* (\langle p_1, \gamma v\rangle,\theta_1) \Rightarrow (\langle p_2, w_2 v\rangle,\theta_1)\Rightarrow^* 	(\langle p'',\epsilon\rangle,\theta'') 
	\end{equation*}
  	Furthermore, 
			if $b=b' \vee b_1 \vee b''=1$, then $b'=1$ or $b_1\vee b''=1$.
		
		\medskip
		For the first case, $b'=1$, then we can have $(\langle p,u \rangle,\theta) \Rightarrow^r (\langle p_1,\epsilon \rangle,\theta_1)$ from (2). Thus, we can obtain that $(\langle p,u \gamma v \rangle,\theta) \Rightarrow^r (\langle p_1,\gamma v \rangle,\theta_1) \Rightarrow^*(\langle p'',\epsilon \rangle,\theta'')$ i.e. $(\langle p,w\rangle, \theta)\Rightarrow^r (\langle p'',\epsilon \rangle,\theta'')$.
		
					\medskip
			The second case $b_1\vee b''=1$ i.e. $B(p_1) \vee b_2 \vee b''=1$  implies that $B(p_1)=1$ (that means $p_1 \in G$ and $(\langle p_1,\gamma v \rangle,\theta_1) \Rightarrow^r (\langle p'',\epsilon \rangle,\theta'')$) or $b_2 \vee b''=1$ (that implies  $(\langle p_2, w_2 v\rangle,\theta_1)\Rightarrow^r 	(\langle p'',\epsilon\rangle,\theta'') $ from (6)). Therefore, $	(\langle p,w\rangle,\theta_1)\Rightarrow^r 
			 (\langle p'',\epsilon \rangle,\theta'')$.
			
			 	\medskip
	\item Case $t$ was added by rule $\alpha_2:$ there exist $p_2 \in P$ and $\theta_2\subseteq \Delta \cup \Delta_c$ such that
\begin{equation*}
	\tag{7}
				r=p_1  { \smrule{(\sigma,\sigma')} p_2 \in \Delta_c \cap \theta_2,\theta_2=(\theta_1 \backslash \sigma) \cup \sigma'}
	\end{equation*}
	and the following path in the current automaton ( self-modifying rule won't change the stack) with $r\in\theta_2:$	
	\begin{equation*}
	\tag{8}
	(p_2,\theta_2) {\xrightarrow[i-1]{[\gamma,b_1']}_{T}} (p_0,\theta_0) \xlongrightarrow[i]{[v,b'']}_{T}(p'',\theta''), ~~ b_1=B(p_1) \vee b_1'
	\end{equation*}
	Applying the transition rule, we can get from (7) that
	\begin{equation*}
	\tag{9}
	(\langle p_1,\gamma v\rangle, \theta_1) \Rightarrow (\langle p_2,\gamma v \rangle, \theta_2)
	\end{equation*}
	We can apply the induction hypothesis (on $j$) to (8), and obtain
	\begin{equation*}
	\tag{10}
	(\langle p_2,\gamma v\rangle, \theta_2) \Rightarrow^* (\langle p'',\epsilon \rangle,\theta'')\text{, moreover, if }b_1'\vee b''=1,(\langle p_2,\gamma v\rangle, \theta_2) \Rightarrow^r (\langle p'',\epsilon \rangle,\theta'')
	\end{equation*}
	
	From (2),(9) and (10), we get
	
	\begin{equation*}
	(\langle p,w\rangle, \theta) =(\langle p,u \gamma v\rangle, \theta)\Rightarrow^* (\langle p_1,\gamma v\rangle,\theta_1)\Rightarrow(\langle p_2,\gamma v\rangle,\theta_2)\Rightarrow^*(\langle p'',\epsilon \rangle,\theta'')
		\end{equation*}
		Furthermore, 
			if $b=b' \vee b_1 \vee b''=1$ , then $b'=1$ or $b_1\vee b''=1$.
		
		\medskip
		For the first case, $b'=1$, then we can have $(\langle p,u \rangle,\theta) \Rightarrow^r (\langle p_1,\epsilon \rangle,\theta_1)$ from (2). Thus, we can obtain that $(\langle p,u \gamma v \rangle,\theta) \Rightarrow^r (\langle p_1,\gamma v \rangle,\theta_1) \Rightarrow^* (\langle p'',\epsilon \rangle,\theta'')$ i.e. $(\langle p,w\rangle, \theta)\Rightarrow^r (\langle p'',\epsilon \rangle,\theta'')$. The second case $b_1\vee b''=1$ i.e. $B(p_1) \vee b_1' \vee b''=1$  implies that $B(p_1)=1$ (that means $p_1 \in G$ and $(\langle p_1,\gamma v \rangle,\theta_1) \Rightarrow^r (\langle p',\epsilon \rangle,\theta')$) or $b_1' \vee b''=1$ (that implies  $(\langle p_2, \gamma v\rangle,\theta_2)\Rightarrow^r 	(\langle p'',\epsilon\rangle,\theta'') $ from (10)) i.e. $	(\langle p,w\rangle,\theta_1)\Rightarrow^r 
			 (\langle p',\epsilon \rangle,\theta')$. Therefore, we can get that if $b=1$, then  $	(\langle p,w\rangle,\theta_1)\Rightarrow^r 
			 (\langle p'',\epsilon \rangle,\theta'')$.

\end{enumerate}


\end{Proof}

 \begin{lemma}
 \label{prelemma2}
	 If there is a labelled configuration $[(\langle p,w \rangle,\theta),b]$ such that $(\langle p,w \rangle, \theta) \Rightarrow^* (\langle p', \epsilon \rangle,\theta')$, then   there is a path $(p,\theta)\by{[w,b]}_{T}   {(p',\theta')}$ in $\mathcal{A}_{pre^*}\big((\langle p',\epsilon \rangle ,\theta')\big)$.
	 Moreover, if  $(\langle p,w \rangle,\theta) \Rightarrow^r (\langle p',\epsilon \rangle,\theta')$, then $b=1.$  
	
\end{lemma}
\begin{Proof}
	Assume $(\langle p,w \rangle,\theta) \stackrel{i}{\Rightarrow} (\langle p',\epsilon \rangle,\theta')$. We proceed by induction on $i$.
	
	\medskip
\noindent
 \textbf{Basis.} $i=0$. Then $\theta =\theta',p'=p$ and $w=\epsilon$.  {Initially, we have that $Q=\{ (p',\theta')\}$, therefore, by the definition of $\rightarrow_{T}$, we have  $(p',\theta')\xrightarrow{\epsilon}_{T}(p',\theta')$. We cannot have $(\langle p',\epsilon \rangle,\theta') \Rightarrow^r (\langle p',\epsilon \rangle,\theta') $ in 0-step.}
	
	\medskip
\noindent
	\textbf{Step.} $i>0$. Then there exists a configuration $(\langle p'', u \rangle,\theta'')$ such that 
\begin{equation*}
(\langle p, w\rangle,\theta) \Rightarrow (\langle p'',u\rangle, \theta'') \stackrel{i-1}{\Rightarrow} (\langle p', \epsilon \rangle,\theta')
\end{equation*}
We apply the induction hypothesis to $(\langle p'',u\rangle, \theta'') \stackrel{i-1}{\Rightarrow} (\langle p', \epsilon \rangle,\theta')$, and obtain  {that there exists in $\mathcal{A}_{pre^*}\big((\langle p',\epsilon \rangle,\theta') \big)$ a  path } $(p'',\theta'')\by{[u,b'']}_{T}(p',\theta')$. If $(\langle p'',u\rangle, \theta'') \Rightarrow^r (\langle p', \epsilon \rangle,\theta')$, $b''=1$.

Let $(p_0,\theta_0)$  be a state of $\mathcal{A}_{pre^*}$. Let $w_1,u_1\in \Gamma^*,\gamma\in \Gamma,b_0'',b_1'' \in \{0,1 \}$ be such that $w=\gamma  w_1$, $u=u_1  w_1$, $b''=b_0'' \vee b_1''$ and
\begin{equation}
	(p'',\theta''){\by{[u_1,b_0'']}}_{T}(p_0,\theta_0) {\by{[w_1,b_1'']}}_{T}(p',\theta')
\end{equation}
There are two cases depending on which rule is applied to get  $(\langle p, w\rangle,\theta) \Rightarrow (\langle p'',u\rangle, \theta'')$.
	
\begin{enumerate}
	\item Case $(\langle p, w\rangle,\theta) \Rightarrow (\langle p'',u\rangle, \theta'') $ is  obtained by a rule of the form: $\langle p,\gamma \rangle \hookrightarrow \langle p'',u_1 \rangle \in \Delta$.  In this case, $\theta''=\theta.$
	   By the saturation rule $\alpha_1$, we have 
	  
	  \begin{equation}
	  	(p,\theta'') {\by{[\gamma,b_0]}}_{T}(p_0,\theta_0), ~b_0=B(p) \vee b_0''
	  \end{equation}
	  
	 Putting (1) and (2) together, we can obtain that 
	 \begin{equation}
	 	\pi=(p,\theta'') {\by{[\gamma,b_0]}}_{T}(p_0,\theta_0)  {\by{[w_1,b_1'']}}_{T}(p',\theta')
	 \end{equation}
	 
	 Thus, $(p,\theta'')\by{[\gamma w_1,b_0\vee b_1'']}_{T}(p',\theta')$ i.e. $(p,\theta)\by{[w,b]}_{T} (p',\theta')$ where $b=b_0 \vee b_1''$. 
	 
\item Case $(\langle p, w\rangle,\theta) \Rightarrow (\langle p'',u\rangle, \theta'') $ is obtained by a rule of the form  { $p \smrule{(\sigma,\sigma')} p'' \in \Delta_c$} i.e $\theta'' \neq \theta.$  In this case, $u_1=\gamma$. By the saturation rule $\beta_2$, we obtain that
	 	\begin{equation}
	 		(p,\theta){\by{[\gamma,b_0]}}_{T}(p_0,\theta_0) \text{ where }\theta''=\theta \backslash \{ r_1\} \cup \{ r_2\}, b_0=B(p)\vee b_0''. 
	 	\end{equation}
	 	
	 	Putting (1) and (4) together, we have the following path
	 	\begin{equation}
	 		(p,\theta){\by{[\gamma,b_0]}}_{T} (p_0,\theta_0){\by{[w_1,b_1'']}}_{T}(p',\theta')\text{ i.e. } (p,\theta){\by{[w,b]}}_{T}(p',\theta') \text{ where } b=b_0 \vee b_1''
	 	\end{equation}
	 	\end{enumerate}
	 	Furthermore, if $(\langle p,w \rangle,\theta)\Rightarrow^r (\langle p',\epsilon \rangle,\theta')$, then $(\langle p, w\rangle,\theta) \Rightarrow^r (\langle p'',u\rangle, \theta'')$  or $(\langle p'',u\rangle, \theta'') \Rightarrow^r (\langle p', \epsilon \rangle,\theta')$.
	 
	 \medskip
	 For the first case, $(\langle p, w\rangle,\theta) \Rightarrow^r (\langle p'',u\rangle, \theta'')$, then $p\in G$ i.e. $B(p)=1$. For the second case, $(\langle p'',u\rangle, \theta'') \Rightarrow^r (\langle p', \epsilon \rangle,\theta')$, we can get $b''=1$ (from induction hypothesis). Thus, $b=b_0 \vee b_1'' =B(p) \vee b_0''\vee b_1''= B(p) \vee b''=1$. Therefore, if $(\langle p,w \rangle,\theta)\Rightarrow^r (\langle p',\epsilon \rangle,\theta')$, then we can obtain $b=1$.

\end{Proof}

\noindent
From these two lemmas, we get:

\begin{theorem}
\label{theorempre}
Let  $[c,b]$ be a labelled configuration.
Then  $[c,b]$ is in $L(\mathcal{A}_{pre^*}\big((\langle p',\epsilon \rangle ,\theta')\big)$ iff  
$c\in pre^*\big((\langle p',\epsilon \rangle ,\theta')\big)$. Moreover, 
  $c\Rightarrow^r(\langle p',\epsilon \rangle ,\theta')$ iff $b=1$. 

\end{theorem}
\begin{Proof}
	Let $[(\langle p, w\rangle, \theta),b]$ be a configuration of $pre^*\big((\langle p',\epsilon \rangle ,\theta')\big))$. Then $(\langle p,w\rangle,\theta) \Rightarrow^*(\langle p', \epsilon\rangle, \theta')$. By Lemma \ref{prelemma1}, we can obtain that there exists a path $(p,\theta) \xrightarrow{[w,b]}_{T}(p',\theta')$ in $\mathcal{A}_{pre^*}\big( (\langle p',\epsilon \rangle,\theta')\big)$. So $[(\langle p, w\rangle, \theta),b]$ is  in $L(\mathcal{A}_{pre^*}\big((\langle p',\epsilon \rangle ,\theta')\big))$. Moreover,  if $(\langle p,w\rangle,\theta) \Rightarrow^r(\langle p', \epsilon\rangle, \theta')$, then $b=1$.

\medskip
\noindent
Conversely, let $[(\langle p, w \rangle,\theta),b]$ be a configuration accepted by $\mathcal{A}_{pre^*}\big((\langle p',\epsilon \rangle ,\theta')\big)$ i.e. there exists a path $(p,\theta) \xrightarrow{[w,b]}_{T}(p',\theta')$  in $\mathcal{A}_{pre^*}\big((\langle p',\epsilon \rangle ,\theta')\big)$. By Lemma \ref{prelemma2}, $(\langle p,w \rangle, \theta)\Rightarrow^* (\langle p',\epsilon \rangle, \theta') $ i.e. $(\langle p,w \rangle, \theta) \in pre^*(L(A))$. Moreover, if $b=1$, $(\langle p,w\rangle,\theta) \Rightarrow^r(\langle p', \epsilon\rangle, \theta')$.

\end{Proof}

 \subsection{Computing the Head Reachability Graph $\mathcal{G}$}

Based on the definition of the  Head Reachability Graph $\mathcal{G}$, and on Theorem \ref{theorempre}, we can compute $\mathcal{G}$ as follows. 
Initially, $\mathcal{G}$ has no edges.

\begin{enumerate}
 \item[$\alpha_1'$:]  if $r_c: p$  {$\smrule{(\sigma,\sigma')}$} $p' \in \Delta_c$, then for every phase $\theta$ such that
$r_c\in\theta$ and every $\gamma\in\Gamma$, we add the edge $((p,\gamma),\theta)\xrightarrow{B(p)}((p',\gamma),\theta_0)$ to  the graph $\mathcal{G}$, where 
$\theta_0=\theta\setminus  {\sigma\cup \sigma'}$.

\item[$\alpha_2'$:] if $r:\langle p,\gamma \rangle \hookrightarrow \langle p_0,\gamma_0 \rangle \in \Delta$, then for every phase $\theta$ such that
$r\in\theta$, we add the edge $((p,\gamma),\theta)\xrightarrow{B(p)}((p_0,\gamma_0),\theta)$ to the graph $\mathcal{G}$.

 	   \item[$\alpha_3'$:] 
if $r: \langle p, \gamma \rangle \hookrightarrow \langle p_0, \gamma_0\gamma' \rangle \in \Delta$, 
then for every phase $\theta$  such that
$r\in\theta$, we add  to  the graph $\mathcal{G}$ the edge $((p,\gamma),\theta)\xrightarrow{B(p)}((p_0,\gamma_0),\theta)$.
Moreover, for  every control point $p'\in P$ and phase  $\theta'$ such that   
$\mathcal{A}_{pre^*}\big((\langle p',\epsilon \rangle ,\theta')\big)$ contains  a transition 
of the form $t=(p_0,\theta)\by{[\gamma_0,b]}_{T}(p',\theta')$, 
we add  to  the graph $\mathcal{G}$ the edge $((p,\gamma),\theta)\xrightarrow{b\vee B(p)}((p',\gamma'),\theta')$.

\end{enumerate}

Items $\alpha_1'$ and $\alpha_2'$ are obvious.   {They respectively correspond to item 1 and item 2 of Definition \ref{graph} (since $B(p)=1$ iff $p\in G$). }
Item $\alpha_3'$ is based on   { Lemma  \ref{lemma1}} and on item 3 of  Definition \ref{graph}. Indeed, it follows from    { Lemma  \ref{lemma1}} that 
$\mathcal{A}_{pre^*}\big((\langle p',\epsilon \rangle ,\theta')\big)$ contains  a transition 
of the form $(p_0,\theta)\by{[\gamma_0,b]}_{T}(p',\theta')$ implies that 
$(\langle p_0,\gamma_0 \rangle,\theta) \Rightarrow^* (\langle p',\epsilon \rangle ,\theta')$, and if $b=1$, then 
$(\langle p_0,\gamma_0 \rangle,\theta) \Rightarrow^r (\langle p',\epsilon \rangle ,\theta')$. Thus, in this case, the edge $((p,\gamma),\theta)\xrightarrow{b\vee B(p)}((p',\gamma'),\theta')$ is added to $\mathcal{G}$ (item 3 of Definition \ref{graph}) since $\langle p,\gamma \rangle \hookrightarrow \langle p_0,\gamma_0\gamma'\rangle \in \Delta$.

\section{Experiments}
\subsection{Our approach vs. standard LTL for PDSs}
We implemented our approach in a tool   and we compared its performance  against the approaches that consist in translating the SM-PDS to an equivalent standard (or symbolic) PDS,
and then applying the standard LTL model checking algorithms  implemented in the PDS  model-checker tool  Moped \cite{Schwoon:2007vs}. 
 All  our  experiments were run on  Ubuntu 16.04 with a 2.7 GHz CPU, 2GB of memory.  
To perform the comparison, we randomly generate several SM-PDSs and LTL formulas  of different sizes.  
The  results   {(CPU Execution time) }are  shown in Table \ref{vsPDS}. \textbf{Column} \emph{Size} is the size of SM-PDS  ($S_1$ for non-modifying  transitions $\Delta$ and $S_2$ for modifying transitions 
$\Delta_c$). 
	\textbf{Column} \emph{LTL} gives  the size of the transitions of the B\"{u}chi automaton generated from the  LTL formula (using the tool LTL2BA\cite{gastin2001fast}).  	 \textbf{Column} \emph{SM-PDS} gives  the cost of our direct algorithm presented in this paper. 
\textbf{Column} \emph{PDS} shows the cost it takes to get the equivalent PDS from the SM-PDS.
	\textbf{Column} {\emph{Result} reports the cost it takes to run the LTL PDS model-checker  Moped \cite{Schwoon:2007vs} for the  PDS we got. 	
\textbf{Column} \emph{Total} is the total cost it takes to translate the SM-PDS into a PDS and then apply the standard LTL model checking algorithm of Moped 
(Total=PDS+Result). { \textbf{Column} \emph{Symbolic PDS}  reports the cost it takes to get the equivalent Symbolic PDS from the SM-PDS. 
\textbf{Column} \emph{$Result_1$} is the cost to run the Symbolic PDS LTL model-checker   Moped.
 \textbf{Column} \emph{$Total_1$} is the total cost it takes to translate the SM-PDS into a symbolic PDS and then apply the standard LTL model checking algorithm of Moped.  
You can see that our direct algorithm (\textbf{Column} \emph{SM-PDS}) is much more efficient than translating the SM-PDS to an equivalent (symbolic) PDS, and then run
the standard LTL model-checker Moped.
{\bf Translating the SM-PDS to a standard PDS may take more than 20 days, whereas our direct algorithm takes only a few seconds.}
Moreover, since the obtained standard (symbolic) PDS is huge, Moped failed to handle several  cases (the time limit that we set for Moped is 20 minutes), 
whereas our tool was able to deal with all the cases in only a few seconds.

 \begin{table}[!hbp]

 \setlength{\tabcolsep}{0.8pt}
 \centering
			\begin{tabular}{|c|c|c|c|c|c|c|c|c|}
			\hline
			\hline
			Size& LTL & SM-PDS & PDS&  Result &Total& Symbolic PDS& $Result_1$&$Total_1$  \\
			\hline
			$S_1:5,S_2:2$&$|\delta|$:15 & \textbf{0.07s} &0.09s &0.01s &0 .10s&0.08s&0.00s&0.08s \\
			\hline
			$S_1:5,S_2:3$&$|\delta|$:8 & \textbf{0.06s} &0.08s& 0.01s &0.09s & 0.09s&0.00s&0.09s\\
				\hline
			$S_1:11,S_2:4$&$|\delta|$:8 & \textbf{0.16s} &0.13s& 0.05s &0.18s & 0.10s&0.00s&0.10s \\
			\hline
			$S_1:5,S_2:3$&$|\delta|$:10 & \textbf{0.06s} &0.15s& 0.01s &0.16s & 0.09s&0.00s&0.09s\\
				\hline

			$S_1:110,S_2:4$&$|\delta|$:8& \textbf{0.34s}&186.10s&0.79s &186.99s & 0.35s&0.00s&0.35s\\
			\hline
			
			$S_1:255,S_2:8$&$|\delta|$:8& \textbf{ 0.39s}&281.02s&0.94s &281.96s &4.82s& 0.05s&4.87s\\
			\hline
			$S_1:255,S_2:8$&$|\delta|$:10& \textbf{0.42s}&281.02s&0.97s &281.99s &4.82s& 0.06s&4.88s\\
			\hline
			$S_1:110,S_2:4$&$|\delta|$:15&\textbf{ 0.28s}&186.10s&1.05s &187.15s & 0.35s&0.06s&0.41s\\
			\hline
			
			$S_1:255,S_2:8$&$|\delta|$:15& \textbf{0.46s}&281.02s&1.92s &282.94s &4.82s& 0.08s&4.90s\\
			\hline
			$S_1:110,S_2:4$&$|\delta|$:20& \textbf{0.37s}&186.10s&1.05s &187.15s & 0.35s&0.06s&0.41s\\
			\hline
			
			$S_1:255,S_2:8$&$|\delta|$:20& \textbf{0.55s}&281.02s&1.97s &282.99s &4.82s& 0.17s&4.99s\\
			\hline
             $S_1:255,S_2:8$&$|\delta|$:25& \textbf{ 0.59s}&281.02s&1.23s &282.99s &4.82s& 0.24s&5.36s\\
			\hline
			$S_1:2059,S_2:7$ &$|\delta|$:8 &\textbf{0.86s } &19525.01s& 20.71s&19545.72s& 20.70s& error&- \\
			\hline
			
			$S_1:2059,S _2:9$ &$|\delta|$:8&\textbf{1.49s} &19784.7s& 79.12s& 19863.32 & 128.12s&error&- \\	
			\hline
			$S_1:2059,S_2:11$ &$|\delta|$:8&\textbf{3.73s} &30011.67s&  168.15s & 30179.82s & 261.07s&error&-\\			
			\hline
			$S_1:2059,S_2:11$ &$|\delta|$:28&\textbf{6.88s }&30011.67s&  169.55s & 30180.22s & 261.07s&error&-\\
			\hline
$S_1:3050,S_2:10$ &$|\delta|$:8&\textbf{5.21s }&39101.57s&killed & - & 438.27s&error&-\\						
			\hline		
			$S_1:3090,S_2:10$ &$|\delta|$:8&\textbf{5.86s} &40083.07s&killed & - & 438.69s &error&- \\
			\hline
			$S_1:3050,S_2:10$ &$|\delta|$:20&\textbf{7.24s} &39101.57s&killed & - & 438.27s&error&-\\						
\hline		
			$S_1:3090,S_2:10$ &$|\delta|$:30&\textbf{8.38s} &40083.07s&killed & - & 438.69s &error&- \\

			\hline
			$S_1:3090,S_2:10$ &$|\delta|$:25&\textbf{8.89s} &40083.07s&killed & - & 438.69s &error&- \\

			\hline
			$S_1:4050,S_2:10$ &$|\delta|$:8&\textbf{9.21s }&81408.91s&  killed & - &699.19s&error&-\\
			\hline	
			$S_1:4050,S_2:10$ &$|\delta|$:28&\textbf{11.64s} &81408.91s&  killed & - &699.19s&error&-\\
			\hline	
			$S_1:4058,S_2:11$ &$|\delta|$:8&\textbf{9.83s} &93843.37s& killed & - & 802.07s&error&-\\
			\hline
			$S_1:4058,S_2:11$ &$|\delta|$:25&\textbf{13.59s }&93843.37s& killed & - & 802.07s&error&-\\
			\hline
			$S_1:5050,S_2:11$ &$|\delta|$:8&\textbf{10.34s} &173943.37s& killed & - &921.16s&error&-\\		
			\hline

			$S_1:5090,S_2:11$ &$|\delta|$:8&\textbf{10.52s} &179993.54s& killed & - & 929.32s&error&- \\
			\hline
			$S_1:5090,S_2:11$ &$|\delta|$:10&\textbf{12.89s }&179993.54s&  killed & - &929.32s&error&-\\	
				\hline
				$S_1:6090,S_2:11$ &$|\delta|$:8&\textbf{13.49s} &190293.64s& killed & -& 1002.73s&error&-\\
				\hline
			$S_1:6090,S_2:11$ &$|\delta|$:10&\textbf{15.81s} &190293.64s& killed & -&1002.73s&error&- \\	
					\hline
					$S_1:6090,S_2:11$ &$|\delta|$:40&\textbf{32.39s }&190293.64s& killed & -& 1002.73s&error&-\\
				\hline

					$S_1:7090,S_2:11$ &$|\delta|$:25&\textbf{39.86s} &198932.32s& killed & - & 1092.28s&error&-\\
					\hline
					$S_1:7090,S_2:11$ &$|\delta|$:30&\textbf{43.24s} &198932.32s& killed & - & 1092.28s&error&-\\
					\hline
					$S_1:9090,S_2:11$ &$|\delta|$:8&\textbf{29.98s} &199987.98s&  killed & - &1128.19s &error&-\\
						\hline
						$S _1:9090,S_2:11$ &$|\delta|$:20&\textbf{45.29s }&199987.98s&  killed & - &1128.19s &error&-\\
						\hline
						$S_1:10050,S_2:12$ &$|\delta|$:8&\textbf{48.53s} &2134587.14s&  killed & - & 1469.28s &error&- \\
						\hline
						$S_1:10050,S_2:12$ &$|\delta|$:25&\textbf{59.69s} &2134587.14s&  killed & - & 1469.28s &error&- \\
						\hline
						$S_1:10050,S_2:12$ &$|\delta|$:30&\textbf{61.42s} &2134587.14s&  kille d & - &1469.28s &error&- \\
						\hline
						$S_1:10150,S_2:12$ &$|\delta|$:35&\textbf{64.17s} &2134633.28s&  killed & -&1469.28s &error&-\\
						\hline
						$S_1:10150,S_2:14$ &$|\delta|$:8&\textbf{58.34s} &2181975.64s& killed & - & 2849.96s&error&-\\
						\hline	
						$S_1:10150,S_2:14$ &$|\delta|$:40&\textbf{82.72s} &2181975.64s& killed & - & 2849.96s &error&-\\
						\hline
						$S_1:10150,S_2:12$ &$|\delta|$:40&\textbf{76.61s} &2134633.28s&  killed & -&1469.28s &error&-\\
						\hline
						$S_1:10150,S_2:16$ &$|\delta|$:45&\textbf{89.83s} &2211008.82s& killed & -&3665.59s &error&-\\
						\hline
						$S_1:10150,S_2:12$ &$|\delta|$:60&\textbf{97.56s} &2134633.28s&  killed & -&1469.28s &error&-\\

 			\hline
						$S_1:10150,S_2:12$ &$|\delta|$:65&\textbf{105.89s} &2134633.28s&  killed & -&1469.28s &error&-\\
						\hline
						$S_1:10150,S_2:16$ &$|\delta|$:65&\textbf{134.45s }&2211008.82s& killed & -&3665.59s &error&-\\
						\hline
						$S_1:10180,S_2:16$ &$|\delta|$:65&\textbf{175.29s} &2134643.52s&  killed & -&3689.83s &error&-\\
						\hline

 				$S_1:10180,S_2:16$ &$|\delta|$:78&\textbf{214.36s} &2134643.52s& killed & -&3689.83s &error&-\\
						\hline

			\hline
		\end{tabular}
		\caption{Our approach vs. standard LTL for PDSs}\label{vsPDS}
	\end{table}

\normalsize
\subsection{Malicious Behavior Detection on Self-Modifying Code}

\subsubsection{Specifying Malicious Behaviors using LTL.}  
As described in \cite{ST13}, several  malicious behaviors can be described  by LTL formulas. We give in what follows three examples of such malicious behaviors and show how they can be described by LTL formulas:

  \medskip
  \noindent \textbf{Registry Key Injecting:} In order to get started at boot
time, many malwares add themselves into the registry key listing. This behavior is typically implemented by first calling the API function  GetModuleFileNameA 
to retrieve the path of the malware's executable file. Then,  the API function RegSetValueExA is called to add the file path into the registry key
listing. This malicious behavior can be described in LTL as follows:

$ \phi_{rk}=\textbf{F} \big(call~GetModuleFileNameA  \wedge \textbf{F}(~call~RegSetValueExA) \big)$

This formula expresses that if a call to the API function GetModuleFileNameA is followed by a call to the  API function RegSetValueExA, then probably a malware is trying to 
add itself into the registry key listing.

 \medskip

\noindent 
 	\textbf{Data-Stealing:} Stealing data from the host is a popular malicious behavior that intend to steal any valuable information 
including passwords, software codes,  bank information, etc. To do this, the malware needs to scan the disk to find the interesting file that he wants to steal.
After finding the file, the malware needs to locate it. To this aim, the malware first calls the API function GetModuleHandleA to get a base address  
 to search for a location of the file.  Then the malware starts looking for the  interesting file  by calling the API function FindFirstFileA. 
Then the API functions CreateFileMappingA and MapViewOfFile are  called to access the file.  Finally, the specific file can be  copied by calling the API function CopyFileA.
Thus, this  data-stealing malicious behavior can be described by the following   LTL formula as follows:

\small 
\noindent
$\phi_{ds}=\textbf{F}(call~GetModuleHandleA~ \wedge \textbf{F} (call~FindFirstFileA 
\wedge 
\textbf{F} ~ (call~CreateFileMappingA~\\
\wedge
\textbf{F}~ 
(call~MapViewofFile
 	 \wedge \textbf{F} ~call~CopyFileA))))$

 \normalsize
 
  \medskip
 
 \noindent
\textbf{Spy-Worm:}
A spy worm is a malware  that  can record  data and send it  using the Socket API functions. 
For example, Keylogger is a spy worm that  can record the keyboard states by calling the API functions GetAsyKeyState  and GetKeyState 
 and send that to the specific server by calling the socket function sendto.
Another   spy worm can also spy on the I/O device rather than the keyboard. For this, it can use the API function  GetRawInputData 
to obtain input from the specified device, and then send this input by calling the socket functions send or  sendto. Thus, this malicious  behavior can be described 
by the following LTL formula:

 \small 
 $\phi_{sw}=\textbf{F} \big((call~GetAsyncKeyState \vee call~GetRawInputData)\wedge \textbf{F}( call~ sendto\vee call~ send)\big)$
 \normalsize

 \medskip
 \noindent
\textbf{Appending virus:}
An appending virus is a virus that inserts a copy of its  code at the end of
the target file. To achieve this, since the real OFFSET of the virus' variables depends on the
size of the infected file, the virus has to first compute its real absolute address
in the memory. To perform  this, the virus has to call the sequence of instructions: {\sf $l_1$: call $f$; $l_2$: ....; $f$: pop eax;}. 
 The instruction {\sf call $f$} will push the return address $l_2$ onto the stack. Then, the pop instruction in $f$ will put
the value of this address into the register eax. Thus, the virus can get its real
absolute address from the register eax. This malicious behavior can be described by the following LTL formula:

\small 
  $\phi_{av}=\bigvee \textbf{F}\Big(call \wedge  \textbf{X} (\mbox{top-of-stack}=a) \wedge \textbf{G} \neg \big(ret \wedge (\mbox{top-of-stack}=a)\big)\Big)$
 \normalsize

\noindent
where the $\bigvee$ is taken over all possible return addresses $a$, and  {top-of-stack}\emph{=a} is a predicate that indicates that the 
top of the stack is $a$.
The subformula  $call \wedge  \textbf{X} (\mbox{top-of-stack}=a)$  means that there exists a procedure
call having $a$ as return address. Indeed,  when a procedure call is made, the program
 pushes its corresponding return address $a$ to the stack. Thus, at the next
step,  $a$ will be  on the top of the stack.  Therefore, the  formula above 
expresses that there exists a  procedure call having $ a$ as  return address, such that there is no  $ret$ instruction which will return to $a$.

\begin{sidewaystable*}[!bp]
 		   \begin{tabular}{|c|c|c|c|c||c|c|c|c||c|c|c|c|} 
 				\hline
 				\hline
 				Example & Size & LTL  & Multiple $pre^*$&  Example & Size&LTL & Multiple $pre^*$ &Example & Size & LTL & Multiple $pre^*$ \\
 				\hline 
 Tanatos.b&12315&16.261s & 46.635s&Netsky.c&45&0.002s & 0.092s &Win32.Happy&23&0.042s&0.075s\\ 			
 					
 				Netsky.a&45&0.047s  & 0.085s &Mydoom.c& 155 & 0.014s & 0.206s&MyDoom-N&16980&30.231s&98.418s \\
 	 			Mydoom.y&26902&12.462s & 102.559s &Mydoom.j& 22355 & 11.262s & 111.617s&klez-N&6281&3.252s&78.419s \\
 			
 			  klez.c&30&0.039s & 0.088s&  Mydoom.v & 5965 & 3.971s & 83.988s &Netsky.b&45&0.057s & 0.183s \\
 			   	
 			   	Repah.b&221&2.428s & 8.852s & Gibe.b&5358&4.229s&17.239s &Magistr.b&4670&3.699s & 93.818s\\
 			   	
 			   	Netsky.d&45&0.083s & 0.123s &  Ardurk.d & 1913 & 0.482s &3.212s &klez.f&27&0.054s & 4.518s\\
 			   	
 			   	Kelino.l&495&0.326s & 5.468s& Kipis.t & 20378 & 23.345s &48.689s &klez.d&31&0.085s & 0.291s\\
 			   			
 			   			Kelino.g & 470 & 0.672s & 3.446s &Plage.b & 395 & 0.291s &3.138s&Urbe.a&123&0.376s & 2.981s \\
 			   				
 			   				klez.e&27&0.094s & 0.482s &Magistr.b&4670&3.987s & 53.235s &Magistr.a.poly&36989&49.863s&159.195s\\

 				Adon.1703&37&0.358s& 0.884s&Adon.1559&37&0.255s & 4.088s& Spam.Tedroo.AB & 487 & 0.924s &4.894s\\
 				Akez&273& 0.136s & 1.863s&Alcaul.d&845&0.165s & 0.392s& Alaul.c & 355 & 0.109s &5.757s\\
 				Haharin.A &210&1.462s & 4.318s&fsAutoB.F026&245&1.698s & 4.503s& Haharin.dr& 235 & 1.558s &4.312s\\

 				 LdPinch.BX.DLL&2010&6.965s&8.128s&LdPinch.fmye&1845&6.194s &9.232s &LdPinch.Win32.5558&2015&6.907s&8.981s\\
 				  			  LdPinch-15&580&1.008s&3.957s&LdPinch.e&578&1.185s&3.392s &Win32/Toga!rfn&590&2.023s&3.978s\\
 				  			  Tanatos.b&12315&16.261s & 46.635s&Netsky.c&45&0.002s & 0.092s &Win32.Happy&23&0.042s&0.075s\\ 	 		
 	 			 	
 	 			Netsky.a&45&0.047s  & 0.085s &Mydoom.c& 155 & 0.014s & 0.206s&MyDoom-N&16980&30.231s&98.418s \\
 	 			Mydoom.y&26902&12.462s & 102.559s &Mydoom.j& 22355 & 11.262s & 111.617s&klez-N&6281&3.252s&78.419s \\
 	 		
 	 		   klez.c&30&0.039s & 0.088s&  Mydoom.v & 5965 & 3.971s & 83.988s &Netsky.b&45&0.057s & 0.183s \\
 	 		    	
 	 		    	Repah.b&221&2.428s & 8.852s & Gibe.b&5358&4.229s&17.239s &Magistr.b&4670&3.699s & 93.818s\\
 	 		    	
 	 		    	Netsky.d&45&0.083s & 0.123s &  Ardurk.d & 1913 & 0.482s &3.212s &klez.f&27&0.054s & 4.518s\\
 	 		    	
 	 		    	Kelino.l&495&0.326s & 5.468s& Kipis.t & 20378 & 23.345s &48.689s &klez.d&31&0.085s & 0.291s\\
 	 		    			
 	 		    			Kelino.g & 470 & 0.672s & 3.446s &Plage.b & 395 & 0.291s &3.138s&Urbe.a&123&0.376s & 2.981s \\
 	 		    				
 	 		    				klez.e&27&0.094s & 0.482s &Magistr.b&4670&3.987s & 53.235s &Magistr.a.poly&36989&49.863s&159.195s\\
 	 		    			   Mydoom-EG[Trj]&230&0.242s & 6.172s &Email.W32!c&220&0.249s & 5.946s &W32.Mydoom.L&235&0.288s&6.452s\\
 	 		        	Mydoom.5&228&0.307s & 8.163s &Mydoom.cjdz5239&225&0.392s & 9.968s &Mydoom.DN.worm&220&0.299s&8.928s\\
 	 		         	Mydoom.R&230&0.322s & 9.086s &Win32.Mydoom&235&0.296s & 7.985s &Mydoom.o@MM!zip&235&0.403s&10.323s\\	
 	 		         	Mydoom.M@mm&5965&5.633s & 108.129s &MyDoom.54464&5935&5.939s & 94.026s &MyDoom.N&5970&6.152s&86.468s\\
 	 		   	Sramota.avf&240&0.383s & 2.691s & Mydoom&238&0.278 & 2.749s &Win32.Mydoom.288&248&0.410s&2.983s\\
 	 		   	 	Win32.Runouce&51678&92.692s & 248.146s &Win32.Chur.A&51895&98.161s & 298.047s &Win32.CNHacker&51095&94.952s&245.452s\\

 	 	 	Win32.Skybag&4180&6.891s & 13.739s &Skybag.A&4310&6.205s  & 15.452s &Netsky.ah@MM&4480&6.991s&16.018s\\
		
 	 			\hline
 	 	Adon.1703&37&0.358s& 0.884s&Adon.1559&37&0.255s & 4.088s& Spam.Tedroo.AB & 487 & 0.924s &4.894s\\
 	 			Akez&273& 0.136s & 1.863s&Alcaul.d&845&0.165s & 0.392s& Alaul.c & 355 & 0.109s &5.757s\\
 	 			Haharin.A&210&1.462s & 4.318s&fsAutoB.F026&245&1.698s & 4.503s& Haharin.dr& 235 & 1.558s &4.312s\\
 	 			
 	 			\hline
 	  LdPinch.BX.DLL&2010&6.965s&8.128s&LdPinch.fmye&1845&6.194s &9.232s &LdPinch..5558&2015&6.907s&8.981s\\
 	 			   	 		    	LdPinch-15&580&1.008s&3.957s&LdPinch.e&578&1.185s&3.392s &Win32/Toga!rfn&590&2.023s&3.978s\\
 	 	LdPinch.by & 970 & 4.092s &11.327s& Generic.2026199&433&2.402s&9.614s &LdPinch.arr & 1250 & 1.848s  & 9.986s \\
 	 	 LdPnch-Fam&195&1.440s&4.097s&Troj.LdPinch.er&205&2.529s&6.154s &LdPinch.Gen.3&210&1.482s&4.973s\\
 	 	 	Androm & 95 & 0.028s &0.192s &  Ardurk.d & 1913 & 3.679s &5.588s &Generic.12861&30183&72.264s&224.809s\\
  Jorik&837&4.159s&11.733s& Bugbear-B&9278&17.737s &52.549s &Tanatos.O&9284&21.481s&79.773s \\

\hline	 
 						\end{tabular}
 			\normalsize
 			\caption{Multiple $pre^*$ v.s. our direct LTL model-checking algorithm}\label{pre*} 		\end{sidewaystable*}

Note that this formula uses predicates that indicate that  the  top of the stack is $a$. Our techniques work for this case as well: it suffices to encode the top of the stack in the 
control points of the SM-PDS. Our implementation works for this case as well and can handle appending viruses.

\vspace{-.3cm}
\begin{sidewaystable*}[!bp]
  \small
 	 	 \begin{tabular}{|c|c|c|c||c|c|c|c||c|c|c|c|} 
 	 			\hline
 	 			\hline
 	 			 Example & Size & Result & $cost$ &  Example & Size&Result & $cost$ &Example & Size & Result & $cost$ \\
 	 			\hline 
 	 			Tanatos.b&12315&Yes & 16.261s&Netsky.c&45&Yes & 0.002s &Win32.Happy&23&Yes&0.042s\\ 	 		
 	 			 	
 	 			Netsky.a&45&Yes & 0.047s &Mydoom.c& 155 & Yes & 0.014s&MyDoom-N&16980&Yes&30.231s \\
 	 			Mydoom.y&26902&Yes & 12.462s &Mydoom.j& 22355 & Yes & 11.262s&klez-N&6281&Yes&3.252s \\
 	 		
 	 		   klez.c&30&Yes & 0.039s&  Mydoom.v & 5965 & Yes & 3.971s &Netsky.b&45&Yes & 0.057s \\
 	 		    	
 	 		    Repah.b&221&Yes & 2.428s & Gibe.b&5358&Yes&4.229s &Magistr.b&4670&Yes & 3.699s\\
 	 		    	
 	 		    Netsky.d&45&Yes & 0.083s &  Ardurk.d & 1913 & Yes &0.482s &klez.f&27&Yes & 0.054s\\
 	 		    	
 	 		    Kelino.l&495&Yes & 0.326s& Kipis.t & 20378 & Yes &25.345s &klez.d&31&Yes & 0.085s\\
 	 		    			
 	 		    		Kelino.g & 470 & Yes & 0.672s &Plage.b & 395 & Yes &0.291s&Urbe.a&123&Yes & 0.376s \\
 	 		    				
 	 		    			klez.e&27&Yes & 0.094s &Magistr.b&4670&Yes & 3.987s &Magistr.a.poly&36989&Yes&49.863s\\
 	 		   	Mydoom.M@mm&5965&Yes & 5.633s &MyDoom.54464&5935&Yes & 5.939s &MyDoom.N!worm&5970&Yes&6.152s\\
 	 		   	Win32.Runouce&51678&Yes & 92.692s &Win32.Chur.A&51895&Yes & 98.161s &Win32.CNHacker.C&51095&Yes&94.952s\\
 	 		     	Win32.Mydoom!O&215&Yes & 0.481s &Mydoom.o@MM!zip&257&Yes & 0.298s &W.Mydoom.kZ2L&228&Yes&0.729s\\
 	 		      	Mydoom-EG [Trj]&230&Yes & 0.242s &Email.Worm.W32!c&220&Yes & 0.249s &W32.Mydoom.L&235&Yes&0.288s\\
 	 		       	Worm.Mydoom-5&228&Yes & 0.307s &Mydoom.CJDZ-5239&225&Yes & 0.392s &Mydoom.DN.worm&220&Yes&0.299s\\

 	 		        	Win32.Mydoom.R&230&Yes & 0.322s &Win32.Mydoom.dlnpqi&235&Yes & 0.296s &Mydoom.o@MM!zip&235&Yes&0.403s\\
 	 		       	Sramota.avf&240&Yes & 0.383s &BehavesLike.Mydoom&238&Yes & 0.278s &Win32.Mydoom.288&248&Yes&0.410s\\
 	 		        	Mydoom.ACQ&19210&Yes & 39.662s &Mydoom.ba&19423&Yes & 38.269s &Mydoom.ftde&19495&Yes&39.583s\\
 	 		          	Worm.Anarxy&210&Yes & 1.913s &Malware!15bf&220&Yes & 2.017s &Anar.A.2&140&Yes&1.993s\\
 	 		           	Win32.Anar.a&215&Yes & 1.631s &nar.24576&240&Yes & 2.738s &Worm-email.Anar.S&155&Yes&2.093s\\
           	HLLW.NewApt&4230&Yes & 6.954s &Win32.Worm.km&4405&Yes & 7.396s &Newapt.Efbh&4550&Yes&7.254s\\
            	NewApt!generic&4815&Yes & 9.002s &NewApt.A@mm&4485&Yes & 8.159s &Newapt.Win32.1&4155&Yes&7.885s\\
             	W32.W.Newapt.A!&5015&Yes & 8.925s &Worm.Mail.NewApt.a&51550&Yes & 9.083s &malicious.154966&5155&Yes&9.291s\\
              Win32.Yanz&2250&Yes & 4.357s &Yanzi.QTQX-0894&2120&Yes & 4.109s &Win32.Yanz.a&2410&Yes&4.465s\\
               	Win32.Skybag&4180&Yes & 6.891s &Skybag.A&4310&Yes & 6.205s &Netsky.ah@MM&4480&Yes&6.991s\\
                	Skybag.b&4955&Yes & 6.892s &Worm.Skybag-1&4820&Yes & 7.119s &Win32.Agent.R&4490&Yes&7.898s\\
                  	Skybag [Wrm]&4985&Yes & 7.482s &Skybag.Dvgb&4830&Yes & 7.564s &Netsky.CI.worm&4550&Yes&7.180s\\

 	 		    			 	 				\hline
 	 		    Agent.xpro & 533 & Yes & 0.352s&Vilsel.lhb&15036&Yes & 4.972s&Generic.2026199&433&Yes&3.489s\\ 	 		
 	 			 	
 	 			Vilsel.lhb&15036&Yes & 26.962s  &Generic.DF&5358&Yes&7.821s&LdPinch.aoq & 7695 & Yes &6.290s \\
 	 		
 	 		    Jorik&837&Yes&4.159s& Bugbear-B&9278&Yes&17.737s &Tanatos.O&9284&Yes&21.481s \\
 	 		    	
 	 		    	Gen.2&1510&Yes&5.632s & Gibe.b&5358&Yes&9.615s &Generic26.AXCN&837&Yes&3.792s\\
 	 		    	
 	 		    	Androm & 95 & Yes &0.028s &  Ardurk.d & 1913 & Yes &3.679s &Generic.12861&30183&Yes&72.264s\\
 	 		    	
 	 		    	LdPinch.by & 970 & Yes &4.092s& Generic.2026199&433&Yes&2.402s &LdPinch.arr & 1250 & Yes  & 1.848s \\
 	 		    			
 	 		    			Generic.12861&30183&Yes&88.294s &Generic.18017273 & 267 & Yes & 0.192s&LdPinch.mg & 5957 & Yes & 9.297s \\
 	 		    				
 	 		    				Script.489524&522&Yes&1.458s&Generic.DF&5358&Yes&8.291s &Zafi&433&Yes&1.028s\\
 	 		    	GenericKD4047614&3495&Yes&4.646s&Win32.Agent.es&3500&Yes&6.083s &W32.HfsAutoB.&3398&Yes&5.092s\\
 	 		    	Trojan.Sivis-1&5351&Yes&7.029s&Win32.Siggen.28&5440&Yes&6.998s &Trojan/Cosmu.isk.&5345&Yes&6.273s\\
 	 		    	Trojan.17482-4&381&Yes&1.495s&Delphi.Gen&375&Yes&1.948s &Trojan.b5ac.&370&Yes&2.089s\\
 	 		    	Delfobfus&798&Yes&3.909s&Troj.Undef&790&Yes&4.068s &Trojan-Ransom.&805&Yes&5.119s\\
 	 		    	LDPinch.400&1783&Yes&4.893s&PSW.LdPinch.plt&1808&Yes&5.088s &PSW.Pinch.1&1905&Yes&5.757s\\
 	 		    	LdPinch.BX.DLL&2010&Yes&6.965s&LdPinch.fmye&1845&Yes&6.194s &LdPinch.Win32.5558&2015&Yes&6.907s\\
 	 		    	TrojanSpy.Lydra.a&3450&Yes&8.289s&Trojan.StartPage&2985&Yes&5.982s &PSWTroj.LdPinch.au&2985&Yes&6.198s\\
 	 		    	LdPinch-21&3180&Yes&6.917s&LdPinch-R&3025&Yes&7.005s &LdPinch.Gen.2&2990&Yes&6.992s\\
 	 		    	Graftor.46303&3230&Yes&5.898s&LdPinch-AIH [Trj]&3010&Yes&6.095s &Win32.Heur.k&2970&Yes&5.950s\\
 	 		    	LdPinch-15&580&Yes&1.008s&LdPinch.e&578&Yes&1.185s &Win32/Toga!rfn&590&Yes&2.023s\\
 	 		    	PSW.LdPinch.mj&595&Yes&1.078s&Gaobot.DIH.worm&590&Yes&1.482s &LDPinch.DF!tr.pws&588&Yes&1.736s\\
 	 		    	TrojanSpy.Zbot&610&Yes&1.610s&LDPinch.10639&605&Yes&1.185s &SillyProxy.AM&590&Yes&1.882s\\
                 LdPinch.mj!c&590&Yes&4.5345s&LdPinch.H.gen!Eldorado&605&Yes&3.955s &Generic!BT&615&Yes&2.085s\\
                  LdPnch-Fam&195&Yes&1.440s&Troj.LdPinch.er&205&Yes&2.529s &LdPinch.Gen.3&210&Yes&1.482s\\
                   Win32.Malware.wsc&150&Yes&2.843s&malicious.7aa9fd&185&Yes&2.189s &WS.LDPinch.400&195&Yes&1.898s\\

 	 		    			\hline
 	 		    		
 	 			\hline
 	 		 	 		\end{tabular}
 	 		\normalsize
 	 	
 	 	\end{sidewaystable*}
 	 	
 	\begin{sidewaystable*}[!bp]
 	\vspace{1cm}
 	\label{tablecomplete}
 	 \small
 	 	 \begin{tabular}{|c|c|c|c||c|c|c|c||c|c|c|c|} 
 	 	 \hline
 	 			\hline
 	 			 Example & Size & Result & $cost$ &  Example & Size&Result & $cost$ &Example & Size & Result & $cost$ \\
 	 			\hline 
 	 			calculation.exe&9952&No & 18.352s &cisvc.exe &4105&No&3.631s &simple.exe&52&No & 0.001s\\ 	 		
 	 			 	
 	 			shutdown.exe &2529&No & 0.397s&loop.exe &529&No & 9.249s&  cmd.exe &1324&No &13.466s\\
 	 			notepad.exe &10529&No & 24.583s&java.exe &800&No & 15.852s&  java.exe &21324&No &42.373s\\
 	 			sort.exe &8529&No &29.789s&bibDesk.exe &32800&No & 50.279s&  interface.exe &1005&No &8.462s\\
 	 			ipv4.exe &968&No &4.186s&TextWrangler.exe &14675&No & 45.221s&  sogou.exe &45219&No &55.259s\\
 	 			game.exe &34325&No &82.424s&cycle.tex &9014&No & 42.555s&  calender.exe &892&No &35.039s\\
 	 			\hline
 	 			SdBot.zk&3430&Yes & 23.242s  &Virus.Gen & 661 & Yes & 9.437s &AutoRun.PR&240& Yes & 4.181s\\ 	 		
 	 			 	
 	 			Adon.1703&37&Yes& 0.358s&Adon.1559&37&Yes & 0.255s& Spam.Tedroo.AB & 487 & Yes &0.924s\\
 	 			Akez&273&Yes & 0.136s&Alcaul.d&845&Yes & 0.165s& Alaul.c & 355 & Yes &0.109s\\
 	 			Virus.Win32.klk&5235&Yes & 15.863s&Virus.Win32.Agent&5340&Yes & 15.968s& Hoax.Gen & 5455 & Yes &13.569s\\
 	 			eHeur.Virus02&420&Yes & 4.985s&Akez.11255&440&Yes & 3.985s& Akez.Win32.1 & 455 & Yes &4.008s\\
 	 				Weird.10240.C&430&Yes & 3.929s&PEAKEZ.A&450&Yes & 2.998s& Virus.Weird.c & 473 & Yes &3.302s\\
 	 				W95/Kuang&435&Yes & 2.985s&Radar01.Gen&465&Yes & 4.005s& Akez.Win32.5 & 490 & Yes &3.958s\\
 	 				Haharin.A&210&Yes & 1.462s&fsAutoB.F026&245&Yes & 1.698s& Haharin.dr& 235 & Yes &1.558s\\
 	 			\hline
	
 	 		NGVCK1&329&Yes & 0.933s&NGVCK2&455&Yes & 1.109s &NGVCK3&2300&Yes&1.388s\\ 	 		
 	 			 	
 	 			NGVCK4&550&Yes & 1.149s &NGVCK5& 1555 & Yes & 1.825s&NGVCK6&1698&Yes&1.689s \\
 	 			NGVCK7&6902&Yes & 14.524s &NGVCK8& 2355 & Yes & 4.254s&NGVCK9&281&Yes&13.301s \\
 	 		
 	 		    NGVCK10&2980&Yes &9.262s& NGVCK11 & 5965 & Yes & 11.456s &NGVCK12&4529&Yes & 10.094s \\
 	 		    	
 	 		    	NGVCK13&2210&Yes & 8.902s & NGVCK14&5358&Yes&10.294s &NGVCK15&970&Yes & 1.912s\\
 	 		    	
 	 		    	NGVCK16&658&Yes & 0.935s & NGVCK17 & 913 & Yes &1.392s &NGVCK18&90&Yes & 0.094s\\
 	 			NGVCK19&1295&Yes & 6.958s& NGVCK20 & 4378 & Yes &15.449s &NGVCK21&31&Yes & 0.097s\\
 	 		    			
 	 		    			NGVCK22 & 370 & Yes & 0.898s &NGVCK23 & 3955 & Yes &9.498s&NGVCK24&6924&Yes & 11.983s \\
 	 		    				
 	 		    			NGVCK25&8127&Yes & 15.018s &NGVCK26&4970&Yes & 9.982s &NGVCK27&7989&Yes&13.197s\\

 	 				NGVCK28&227&Yes & 0.098s &NGVCK29&960&Yes & 0.692s &NGVCK30&89&Yes&0.088s\\
 	 				NGVCK31&550&Yes & 0.875s &NGVCK32&60&Yes & 0.059s &NGVCK33&65&Yes&0.069s\\
 	 		    			NGVCK34&5990&Yes & 9.848s &NGVCK35&4590&Yes & 10.178s &NGVCK36&825&Yes&2.934s\\

 	 				NGVCK37&80&Yes & 0.998s &NGVCK38&150&Yes & 1.093s &NGVCK39&395&Yes&1.048s\\
 	 		    			NGVCK40&40&Yes & 0.921s &NGVCK41&950&Yes & 0.704s &NGVCK42&8290&Yes&15.085s\\
 	 		    			NGVCK43&6220&Yes & 2.930s &NGVCK44&5215&Yes & 11.006s &NGVCK45&9290&Yes&14.595s\\
 	 		    			NGVCK46&320&Yes & 0.928s &NGVCK47&834&Yes & 2.958s &NGVCK48&9810&Yes&14.696s\\
 	 		    			NGVCK49&12320&Yes & 25.395s &NGVCK50&8810&Yes & 19.969s &NGVCK51&39810&Yes&68.283s\\
 	 		    			NGVCK52&520&Yes & 0.289s &NGVCK53&15&Yes & 0.089s &NGVCK54&8883&Yes&11.393s\\
 	 		    			NGVCK55&12520&Yes & 38.768s &NGVCK56&6218&Yes & 15.489s &NGVCK57&32562&Yes&83.482s\\
 	 		    			NGVCK58&9520&Yes & 23.658s &NGVCK59&818&Yes & 2.592s &NGVCK60&12962&Yes&38.025s\\
 	 		    			NGVCK61&10020&Yes & 24.976s &NGVCK62&8818&Yes & 19.299s &NGVCK63&2068&Yes&3.662s\\
 	 		    			NGVCK64&273&Yes & 1.987s &NGVCK65&5855&Yes & 8.995s &NGVCK66&68&Yes&1.002s\\
 	 		    			NGVCK69&4150&Yes & 8.052s &NGVCK70&9860&Yes & 24.199s &NGVCK71&3240&Yes&7.951s\\
 	 		    			NGVCK72&31&Yes & 0.591s &NGVCK73&549&Yes & 1.052s &NGVCK74&9078&Yes&29.078s\\
 	 		    			NGVCK75&90&Yes & 1.002s &NGVCK76&5890&Yes & 10.128s &NGVCK77&1958&Yes&9.559s\\
 	 				NGVCK78&33468&Yes & 75.098s &NGVCK79&4735&Yes & 10.980s &NGVCK80&45273&Yes&82.396s\\
 	 		    			NGVCK66&777&Yes & 0.198s &NGVCK67&895&Yes & 0.223s &NGVCK81&6939&Yes&2.726s\\
 	 		    			NGVCK82&2931&Yes & 0.463s &NGVCK83&8759&Yes & 10.316s &NGVCK84&34563&Yes&53.244s\\
 	 		    			NGVCK85&19024&Yes & 29.220s &NGVCK86&1026&Yes & 0.572s &NGVCK87&7929&Yes&5.671s\\
 	 		    			NGVCK88&6126&Yes & 8.682s &NGVCK89&580&Yes & 2.036s &NGVCK90&27843&Yes&17.353s\\
 	 		    			NGVCK91&20&Yes & 0.001s &NGVCK92&59&Yes & 0.903s &NGVCK93&98&Yes&0.021s\\
 	 		    			NGVCK94&150&Yes & 0.146s &NGVCK95&1679&Yes & 0.294s &NGVCK96&6299&Yes&5.196s\\
 	 		    			NGVCK97&4496&Yes & 5.272s &NGVCK98&428&Yes & 0.329s &NGVCK99&158&Yes&1.153s\\
 	 		    			NGVCK100&895&Yes & 0.961s &NGVCK101&745&Yes & 1.117s &NGVCK102&704&Yes&0.269s\\
 	 		    			NGVCK103&86&Yes & 0.282s &NGVCK104&145&Yes & 0.998s &NGVCK105&24124&Yes&68.816s\\

 	 		    		 	 		    		\hline
 	 	 \end{tabular}
 	 	\caption{ Experimental Results}\label{appendix}
 	 	 \end{sidewaystable*}

\subsubsection{Applying our tool for malware detection.}

We applied our tool to detect several malwares. We use the unpack tool unpacker \cite{unpacker} to handle  packers like UPX, and 
we use Jakstab \cite{jakstab} as  disassembler. 
We consider  160 malwares from the malware library VirusShare \cite{vxShare}, 
184 malwares from the malware library MalShare \cite{malShare}, 288 email-worms from VX heaven \cite{vh} and 260 new malwares generated by NGVCK, one of the best malware generators. We also choose 19 benign samples from Windows XP system.   We consider self-modifying versions of these programs. In these versions, the malicious behaviors 
are unreachable if the semantics of the self-modifying instructions are not taken into account, i.e.,  if the  self-modifying instructions are considered as ``standard'' 
instructions that do not modify the code, then the malicious behaviors cannot be reached. To check this, we model such programs in two ways:
\vspace{-.2cm}
\begin{enumerate}
\item   First, we take into account the self-modifying instructions and model these programs using SM-PDSs as described in Section \ref{model}. Then, we check whether these 
SM-PDSs satisfy  at least one of the malicious LTL formulas presented above.  If yes, the program is declared as malicious, if not, it is declared as benign.
 Our tool was able to detect all the 892 self-modifying malwares as malicious, 
and to determine that benign programs are benign.  We report  in Table  \ref{appendix}  the   results we obtained.
\textbf{Column} \emph{Size} is the number of control locations,
 \textbf{Column} \emph{Result} gives  the result of our algorithm: {\bf Yes} means malicious and {\bf No} means benign; 
 and \textbf{Column} \emph{cost} gives  the cost to apply our LTL model-checker to check one of the LTL properties described above.

\item  Second, we abstract away the self-modifying instructions and proceed as if these instructions were not self-modifying. In this case, we translate 
the binary codes to standard pushdown systems as described in \cite{SongT12}. By using PDSs as models, none of the malwares that we consider was detected as malicious, whereas,
as reported in Table \ref{appendix}, using self-modifying PDSs as models, and applying our LTL model-checking algorithm allowed to detect all the 892 malwares that we considered.
 \end{enumerate}

Note that checking the formulas $\phi_{rk}$, $\phi_{ds}$, and $\phi_{sw}$ could be done using multiple $pre^*$ queries on SM-PDSs using the $pre^*$  algorithm of \cite{touili2017reachability}. 
However, this would be less efficient than performing our direct LTL model-checking algorithm, as shown in Table \ref{pre*},
 where \textbf{Column} \emph{Size} gives 
 the number of control locations,
 \textbf{Column} \emph{LTL} gives  the time of applying our LTL model-checking algorithm; 
 and \textbf{Column} \emph{Multiple $pre^*$} gives  the cost of applying multiple $pre^*$ on SM-PDSs to check the properties  $\phi_{rk}$, $\phi_{ds}$, and $\phi_{sw}$.
It can be seen that applying our {\em direct} LTL model checking algortihm is more efficient.
Furthermore, the appending virus formula $\phi_{av}$  cannot be solved using  multiple $pre^*$ queries. Our direct LTL model-checking algorithm is needed in this case.
Note that some of the malwares we considered in our experiments are appending viruses. Thus, our algorithm and our implementation are crucial to be able to detect these malwares.

\begin{table}[]
\hspace{-1.5cm}
\label{anti}
\begin{tabular}{|c|c|c|c|c|c|c|c|c|c|c|c|c|}
\hline 
 {\bf our tool}& McAfee&  Norman&BitDefender&Kinsoft&Avira&eScan&Kaspersky&Qihoo360&Baidu&Avast&Symantec\\
\hline
 {\bf	100\%}& 24.8\%&19.5\%&31.2\%&9.7\%&34.1\% & 21.9\% & 53.1\% &51.7\% &1.4\%&68.3\% &82.4\%\\
 	 
 \hline
\end{tabular}	
\caption{Detection rate: Our tool vs. well known  antiviruses}\label{antivirus}
\end{table}

 \subsubsection{Comparison with well-known antiviruses.}
We  compare our tool against well-known and widely used antiviruses.
 Since known antiviruses update their signature database as soon as a new malware is known, in order to have a fair comparision with these antiviruses, we need to consider new malwares. We  use the sophisticated malware generator NGVCK available at VX Heavens \cite{vh} to generate 205 malwares.
 We obfuscate these malwares with self-modifying code, and we fed them to our tool 
and to well known antiviruses such as  BitDefender, Kinsoft, Avira, eScan, Kaspersky, Qihoo-360, Baidu, Avast, and Symantec. Our tool was able to detect all these programs as malicious, whereas none of the well-known antiviruses was able to detect all these malwares. Table \ref{antivirus} reports the detection rates of our tool and the well-known anti-viruses.

\bibliographystyle{plain}
\bibliography{paper.bib}

\end{document}